\pdfoutput=1

\documentclass[12pt]{article}
\usepackage{amsfonts}
\usepackage{amsmath}
\usepackage{amssymb}
\usepackage{bigints}
\usepackage{booktabs}
\usepackage[nosort]{cite}
\usepackage{color}
\usepackage{dsfont}
\usepackage{float}
\usepackage{framed}
\usepackage{graphicx}
\usepackage{indentfirst}
\usepackage{mathrsfs}
\usepackage{multirow}
\usepackage{pdflscape}
\usepackage{setspace}
\usepackage{subdepth}
\usepackage{subfig}
\usepackage{titlesec}
\usepackage[dotinlabels]{titletoc}
\usepackage{wrapfig}
\usepackage[all]{xy}
\usepackage{young}
\usepackage[vcentermath]{youngtab}
\usepackage{relsize}
\usepackage{stackengine}
\usepackage{datetime}

\usepackage{hyperref}

\numberwithin{equation}{section}

\usepackage{verbatim}

 \newcommand{\reef}[1]{(\ref{#1})}

\newcommand{\be}{\begin{equation}}
\newcommand{\ee}{\end{equation}}

\newcommand{\bml}{\begin{multline}}
\newcommand{\emll}{\end{multline}}

\newcommand{\nn}{\nonumber}

\def\({\left(} \def\){\right)}
\def\[{\left[} \def\]{\right]}

\def\al{\alpha}

\def\mA{\mathcal{A}}

\def\eps{\epsilon}

\def\b{\bar}

\def\lam{\lambda}

\def\Lam{\Lambda}

\def\del{{\partial}}

\newcommand{\G}{\Gamma}

\def\d{\partial}

\newcommand{\la}{\langle}
\newcommand{\ra}{\rangle}

\newcommand{\bea}{\begin{eqnarray}}
\newcommand{\eea}{\end{eqnarray}}

\usepackage[left=2cm,right=2cm,top=2cm, bottom=2cm]{geometry}
\titleformat{\section}{\normalfont\bfseries}{\thesection.}{4pt}{}
\titlespacing{\section}{0pt}{22pt}{6pt}

\titleformat{\subsection}{\normalfont\itshape}{\thesubsection.}{4pt}{}
\titlespacing{\subsection}{0pt}{18pt}{6pt}

\titleformat{\subsubsection}{\normalfont\itshape}{\thesubsubsection.}{4pt}{}
\titlespacing{\subsubsection}{0pt}{16pt}{6pt}

\def\ie{\begin{equation}\begin{aligned}}
\def\fe{\end{aligned}\end{equation}}

\newcommand{\fdfrac}[2]{\mbox{\footnotesize$\displaystyle\frac{#1}{#2}$}}

\def\tilde{\widetilde}
\def\t{\tilde}

\def\bar{\overline}

\def\d{\partial}

\def\1{{\mathds 1}}

\def\mA{\mathcal{A}}

\def\mL{\mathcal{L}}

\DeclareFontShape{OT1}{cmr}{mx}{n}%
    {<->cmr10}{}
\newcommand{\mytitlefont}{\fontseries{mx}\selectfont}
\DeclareMathAlphabet{\titlemath}{OT1}{cmr}{mx}{n}

\begin{document}


\vfill

\begin{titlepage}

\begin{center}

~\\[2cm]

{\fontsize{20pt}{0pt} \mytitlefont Integrability and Renormalization under $T \b T$}

~\\[0.5cm]

{\fontsize{14pt}{0pt} Vladimir Rosenhaus$^1$ and Michael Smolkin$^2$}

~\\[0.1cm]

\it{$^1$ School of Natural Sciences, Institute for Advanced Study}\\ \it{Princeton, NJ 08540, USA}\\[.5cm]

\it{$^2$ The Racah Institute of Physics, The Hebrew University of Jerusalem, \\ Jerusalem 91904, Israel}

~\\[0.8cm]

\end{center}

\noindent 

Smirnov and Zamolodchikov recently introduced a new class of two-dimensional quantum field theories, defined through a differential change of any existing theory by the determinant of the energy-momentum tensor. From  this $T\b T$ flow equation one can find a simple expression for both the energy spectrum and the $S$-matrix of the $T\b T$ deformed theories. Our goal is to find the renormalized Lagrangian of the $T\b T$ deformed theories. In the context of the $T\b T$ deformation of an integrable theory, the deformed theory is also integrable and, correspondingly, the $S$-matrix factorizes into   two-to-two $S$-matrices. One may thus hope to be able to extract the renormalized Lagrangian from the $S$-matrix. We do this explicitly for the $T\b T$ deformation of a free massive scalar, to second order in the deformation parameter. Once one has the renormalized Lagrangian  one can, in principle, compute all other observables, such as  correlation functions. We briefly discuss this, as well as the relation between the renormalized Lagrangian, the $T\b T$ flow equation, and the $S$-matrix. We also mention a more general class of integrability-preserving deformations of a free scalar field theory.

\vspace{2cm}

\vfill

September 9, 2019
\end{titlepage}

\tableofcontents
~\\[.1cm]
\section{Introduction}

The bootstrap is a powerful, nonperturbative, method to study quantum field theory. Rather than starting with a specific theory, one starts with a set of consistency relations for the $S$-matrix that any theory, or any class of theories, must satisfy. Assuming one succeeds in finding a solution, a question arises: what theory, if any, is this the solution of?

In this paper we study this question in the context of integrable two-dimensional field theories, for a specific type of $S$-matrix. Two-dimensional integrable QFTs can be characterized as theories that have no particle production. One can discuss integrability of the classical theory (no particle production at tree level), as well as integrability of the quantum theory; the former does not always imply the latter. In the simplest example of an integrable theory, the sinh-Gordon model, it so happens that the renormalized  Lagrangian takes the same functional form as the classical Lagrangian, and hence quantum integrability follows immediately from classical integrability.

Recently, Smirnov and Zamolodchikov \cite{SZ} introduced a rich new class of integrable two-dimensional theories. They gave both the $S$-matrix for these theories, as well as the classical Lagrangian. Our goal is to find the renormalized Lagrangian. Renormalization of these theories is highly nontrivial. Indeed, from the form of the classical Lagrangian, one would say that these theories are non-renormalizable. If these were  standard (non-integrable) QFTs, this would mean that one lacks  predictive power in the UV: when renormalizing, the finite part of counterterms must be fixed experimentally, and for non-renormalizable theories there are an infinite number of counterterms. However, these theories are integrable, and this gives an infinite number of constraints which, one may hope, uniquely fix all counterterms.

The central equation in the analysis of Smirnov and Zamolodchikov \cite{SZ} is the $T\b T$ flow equation. Starting from any Lagrangian, the $T \b T$ deformation generates a one-parameter family of Lagrangians  satisfying the equation,
\be \label{DiffEI}
\frac{\d \mL(\lam)}{\d \lam} = - 4 \( T_{z z} ^{\lam}\,  T_{\b z \b z}^{\lam} - (T_{z \b z}^{\lam})^2\)~,
\ee
where $\mL (\lam= 0)$ is the original Lagrangian and $T_{\mu \nu}^{\lam}$ are the components of the energy-momentum tensor of the finite $\lam$ theory. The composite operator on the right-hand side is defined via point splitting. It is important to note that both sides of the equation are renormalized and UV finite; this equation is not an RG flow equation.  Rather, it is an equation describing some particular one-parameter family of theories. From the $T\b T$ flow equation, it is simple to derive a differential equation for the energy spectrum as a function of $\lam$, and to derive the $\lam$ dependence of the $S$-matrix. It is, however, not simple to  find the renormalized Lagrangian; this will be our goal. Our approach will be to first solve the classical version of (\ref{DiffEI}). Starting with any theory, one may solve the classical $T \b T$ flow equation perturbatively in $\lam$, using the classical  energy-momentum tensor for  the right-hand side. The result will be the classical Lagrangian of the $T\b T$ deformation of the original Lagrangian. This Lagrangian must then be renormalized. We find an unambiguous renormalized Lagrangian by demanding that it gives the correct $S$-matrix: the one required by the (renormalized) version of (\ref{DiffEI}). 
 We will do this explicitly to one loop order for the $T \b T$ deformation of a free scalar.

The paper is organized as follows. In Sec.~\ref{Sec2} we review  the $T\b T$ deformation, as well as some elementary aspects of integrability that pertain to it. In Sec.~\ref{Sec3} we consider the $T\b T$ deformation of a free  scalar. In Sec.~\ref{Massless} we compute the renormalized Lagrangian to one loop order for the $T\b T$ deformation of a free massless scalar. In Sec.~\ref{Massive} we compute the renormalized Lagrangian to one loop order for the $T\b T$ deformation of a free massive scalar. An interesting result 
is that the renormalized Lagrangian is qualitatively different from the classical Lagrangian. In Sec.~\ref{Sec4} we discuss further aspects of $T\b T$. In Sec.~\ref{STTbarAp} we discuss the relation between the $T\b T$ flow equation and the $S$-matrix. In   Sec.~\ref{Tmunu} we discuss the relation between the renormalized Lagrangian and the $T\b T$ flow equation. In Sec.~\ref{T2pt}  we discuss the relation between the renormalized Lagrangian and correlation functions under $T\b T$. In Sec.~\ref{Integrability} we show that there is a  broad class of deformations of a free theory, going beyond $T\b T$,  that are integrable, at least classically.
 In Sec.~\ref{Sec5} we discuss future directions. In  Appendices  \ref{AppendixG} and \ref{IntEx} we collect some useful integrals.

\section{Integrability} \label{Sec2}

In this section we review some elementary aspects of integrability of two-dimensional quantum field theories  \cite{ZZ}, as well as some relevant aspects of $T\b T$.

Consider the two-to-two $S$-matrix in a field theory with a single particle species, of mass $m$. The energy and momenta of the particles are parametrized by the rapidity $\theta_i$~, 
\be
E(\theta_i) = m \cosh \theta_i~, \ \ \  p(\theta) = m\sinh \theta_i~.
\ee
As a result of two-dimensional kinematics, the ingoing momenta  are the same as the outgoing momenta. The two-to-two $S$-matrix is thus only a function  of the rapidity difference, $\theta = \theta_1 - \theta_2$, and is denoted by $S(\theta)$. 
Correspondingly, the Mandelstam variables, in signature $(+,-)$, are
\be
s = (p_1 + p_2)^2 = 2m^2 (1+ \cosh \theta)~, \ \ \ \ ~ t = 4m^2 - s~, \ \ \ \ \ \ u=0~.
\ee
The $t$ channel corresponds to $\theta \rightarrow i\pi - \theta$, and the $u$ channel corresponds to $\theta = i\pi$. 

The assumption of integrability is that there is no particle production: the $2$ to $n$\, $S$-matrix, for $n>2$ is zero. As a result, unitarity, which ordinarily is the inequality $|S(\theta)|^2 \leq 1$, becomes the equality  $|S(\theta)|^2 = 1$. Combined with crossing symmetry: the symmetry of the $S$-matrix under the interchange of the $s$ and $t$ channels, we have the set of equations,
\be \label{23}
|S(\theta)|^2 = 1~, \ \ \ \text{and} \ \ \ \ S(i\pi - \theta) = S(\theta)~.
\ee
A solution of these equations is the CDD factor, 
\be \label{SGS}
\mathcal{S}_{\al}(\theta)= \frac{\sinh \theta - i \sin \alpha}{\sinh \theta + i \sin \alpha}~,
\ee
for any real $\alpha$. It is easy to see that this is a solution, since $\sinh(i \pi - \theta) = \sinh \theta$, and taking a ratio as above ensures unitarity. A product of $\mathcal{S}_{\al}(\theta)$ over various $\alpha$, $\prod_{\al} \mathcal{S}_{\al}(\theta)$, is  clearly also a solution. A single  factor, $\mathcal{S}_{\al}(\theta)$, is the $S$-matrix for the sinh-Gordon model, where the parameter $\al$ is related to the coupling. 

\subsection{$T\b T$ flow equation}
 An interesting question is which theory, if any, has an $S$-matrix that is some product of the $\mathcal{S}_{\al}(\theta)$. Progress in this direction was recently made by Smirnov and Zamolodchikov \cite{SZ}. They consider an alternative basis of solutions of (\ref{23}), 
\be \label{25}
\mathcal{S}'_s(\theta) = \exp\( i \lam_s m^{2s} \sinh (s\, \theta)\)~,
\ee
where $s$ is an odd integer, and $\lam_s$ is a constant of dimension $-2s$, so that $\lam_s m^{2s}$ is dimensionless. A result of \cite{SZ} is the following: let $S(\theta)$ be the $S$-matrix of some integrable theory. Since the theory is integrable, it has an infinite number of conserved currents, $\b\d T_{s+1} =   \d \Theta_{s-1}$. Suppose one constructs a one-parameter family of theories, depending on the parameter $\lam_s$, that are deformations of this theory, and have a Lagrangian that solves the differential equation, 
\be \label{TsTs}
\frac{\d \mL}{\d \lam_s} = - 4 \( T_{s+1}^{\lam_s}\, \b T_{s+1}^{\lam_s} - \Theta_{s-1}^{\lam_s} \b \Theta_{s-1}^{\lam_s}\)~,
\ee
where  $\lam_s=0$ corresponds to  the original theory, and  $T_{s+1}^{\lam_s}$ and $\Theta_{s-1}^{\lam_s}$ are the conserved currents of the deformed theory. Then, \cite{SZ} argue that the deformed theory is integrable, and moreover, has an $S$-matrix given by $S(\theta)\, \mathcal{S}'_s(\theta)$.

The simplest case (the $T \b T$ deformation) is that of a deformation with $s=1$, in which case the currents are simply components of the stress tensor, $T_{\mu \nu}$. Since every theory has a stress tensor, the $T\b T$ deformation can be performed on any theory, integrable or not. The differential equation for the $\lam$ dependence of the Lagrangian is therefore,
\be \label{DiffE}
\frac{\d \mL}{\d \lam} = - 4 \( T^{\lam}\, \b T^{\lam} - (T_{z \b z}^{\lam})^2\)~.
\ee
Moreover, consider the simplest subcase of this, in which one deforms a free massive scalar (so that the $S$-matrix of the initial theory is the identity), 
\be \label{Lstart}
\mL= 2\d \phi \b \d \phi + V(\phi)~, \ \ \ \ V = \frac{1}{2} m^2 \phi^2~, 
\ee
where $\d \equiv \d_{z}$ is the derivative with respect to the holomorphic coordinate $z= x_1 + i x_2$, and similarly $\b \d \equiv \d_{\b z}$ is the derivative with respect to the antiholomorphic coordinate $\b z = x_1 - i x_2$. 
The $S$-matrix of the $T\b T$ deformation of the free scalar is (\ref{25}) with $s=1$, 
\be \label{STTbar}
S(\theta) = \exp\( i \lam m^2 \sinh \theta\)~.
\ee
The classical Lagrangian for the $T\b T$ deformation of a massless scalar was first found by Cavagli\`a, Negro, Sz\'ecs\'ecnyi, and Tateo \cite{Tateo16}. The  classical Lagrangian for the $T\b T$ deformation of a scalar theory of the form (\ref{Lstart}), with an arbitrary potential $V(\phi)$, was found in \cite{Bonelli:2018kik} and takes the form, 
\be \label{Lag}
\mL = \frac{V}{1-\lam V} + \frac{-1 + \sqrt{1 + 8\, \b \lam\, \d \phi \,\b \d \phi }}{2\b \lam}~, \ \ \ \ \ \b \lam = \lam \( 1- \lam V\)~.
\ee
For $\lam = 0$, this reduces to the starting Lagrangian (\ref{Lstart}). 
One can explicitly verify that the Lagrangian (\ref{Lag}), combined with its classical stress tensor,  is a solution of the $T\b T$ flow equation (\ref{DiffE}).

We now make several comments: 
\begin{enumerate}
\item
An essential aspect of the $T\b T$ deformation is that one knows both how the theory changes, as described by the flow equation (\ref{DiffE}), and how the $S$-matrix changes, by picking up the phase factor (\ref{STTbar}). Let us briefly review how this connection can be understood. 
One may notice that in the massless case, (\ref{Lag}) is the  Nambu-Goto action for a long string in three spacetime dimensions, in static gauge. In work predating the study of $T\b T$ in the form initiated by \cite{SZ},   Dubovsky, Flauger and Gorbenko \cite{Dubovsky:2012wk} found the $S$-matrix for this Nambu-Goto action to be (\ref{STTbar}); see also \cite{Caselle:2013dra}. Their computation did not explicitly use the Lagrangian. Rather, they appealed to the knowledge of the energy spectrum of a string, and then applied the TBA equation, which (for integrable theories) relates the energy spectrum to the $S$-matrix.
 More generally, if one takes the $S$-matrix for any integrable theory and multiplies it by the factor (\ref{STTbar}) \cite{Dubovsky:2013ira}, then one can turn the TBA equation into a differential equation for the energy spectrum as a function of $\lam$ \cite{Tateo16}. At the same time, one can start from the definition of the $T\b T$ deformation (\ref{DiffE}) and compute the energy spectrum, as a function of $\lam$, to find the same equation \cite{Z, SZ, Tateo16}. This establishes, for integrable theories, that the $T\b T$ deformation (\ref{DiffE}) is equivalent to multiplying the $S$-matrix by the factor (\ref{STTbar}). In fact, the $T\b T$ deformation is not unique to integrable theories; one can study the $T \b T$ deformation of any theory. The relation between the flow equation (\ref{DiffE}) and the change in the $S$-matrix in this general case will be discussed in Sec.~\ref{STTbarAp}. 
 
 \item
 In stating that the Lagrangian (\ref{Lag}) is a solution of the $T\b T$ flow equation (\ref{DiffE}), we treated the flow equation as a classical equation. Consequently, (\ref{Lag})  is the classical Lagrangian. In the study of Smirnov and Zamolodchikov \cite{SZ}, and in particular in the computation of the energy spectrum and the $S$-matrix, the $T\b T$ flow equation is treated as a quantum equation, with operators on the right-hand side.~\footnote{More precisely, the composite operator on the right-hand side is defined by point-splitting, and the right-hand side is defined up to a total derivative (which is irrelevant for the change in the action). The fact that via point-splitting one gets a finite operator is nontrivial; this was shown in \cite{Z}, and relies crucially on both two dimensions and $T_{\mu \nu}$ being a conserved current.  } In particular, the Lagrangian on the left-hand side of the $T\b T$ flow equation  is implicitly taken to be the renormalized Lagrangian. 
Our goal in this paper is to find the renormalized Lagrangian: to renormalize, perturbatively in $\lam$, the classical Lagrangian (\ref{Lag}). The counterterms are chosen so that (\ref{STTbar}) is the $S$-matrix of the renormalized Lagrangian. This is what we do in Sec.~\ref{Sec3}, to one loop level.

 \item In finding consistent $S$-matrices through the bootstrap, it is important to not only impose unitarity and crossing, as was done in Eq.~\ref{23}, but also to ensure that the $S$-matrix has correct analytic behavior. The $S$-matrices (\ref{25}) do not have correct analytic behavior - they grow exponentially at large imaginary momenta. Such growth is inconsistent with the behavior of a local quantum field theory. However, rather than just discarding these theories, the approach in much of the $T\b T$ literature is to regard them as quantum field theories coupled to gravity \cite{Dubovsky:2017cnj, Dubovsky:2018bmo, Cardy, Conti:2018tca, Conti:2018jho}.~\footnote{The AdS$_3$ dual of the $T\b T$ deformed theories is correspondingly an interesting question, first studied in \cite{Herman}. At large $N$ one expects the $T\b T$ deformation to correspond to a change in boundary conditions \cite{Guica:2019nzm}. See also \cite{Hartman,Taylor:2018xcy} for AdS studies, \cite{Kutasov, Giribet:2017imm} for related developments on $T\b T$ and string theory in AdS$_3$, and  \cite{Eva} for  studies of dS. Separately, understanding the Hagedorn behavior and partition function of the $T\b T$ theories is an interesting question \cite{SZ, Tateo16, Datta:2018thy, Jiang:2019hxb, Aharony:2018bad}.  }
 The question of if it is correct to interpret the $T\b T$ deformed theories at  finite $\lam$  as gravitational theories is orthogonal to the focus of this paper: we will work perturbatively in $\lam$ around $\lam =0$, and in this regime these theories are quantum field theories. 
   \end{enumerate}

We finish this section with a brief review of the one-loop $S$-matrix in the sinh-Gordon model; this is a good warmup for the one-loop calculation that we will do in Sec.~\ref{Sec3}.

\subsection{Sinh-Gordon model} \label{Sinh}

As we mentioned, the theory that gives the  CDD factor $S$-matrix in (\ref{SGS}) is the sinh-Gordon model,
\be
\mL = \frac{1}{2} (\d \phi)^2  + \frac{m^2}{b^2} \( \cosh( b \phi) - 1\)~,  \ \ \ \ \ \ \alpha = \frac{\pi b^2}{8\pi + b^2}~, 
\ee
in which the coupling $b$ is related to the free parameter $\alpha$ in the $S$-matrix. For imaginary $b$, this is the sine-Gordon model.  In this section we briefly recall some properties of the sinh-Gordon model, as originally discussed in \cite{Arefeva:1974bk}. %

It is instructive to expand the Lagrangian perturbatively in $b$, 
\be \label{Lexp}
\mL =  \frac{1}{2} (\d \phi)^2 + \frac{1}{2} m^2 \phi^2 + \frac{m^2 b^2}{4!} \phi^4 +  \frac{m^2 b^4}{6!} \phi^6 + \ldots~,
\ee
and  to verify, to a few order in $b^2$, that ({\ref{SGS}) is in fact the correct $S$-matrix. 
An immediate question that arises is that (\ref{Lexp}) is the bare Lagrangian; presumably, we will encounter UV divergences and will need to renormalize the theory. In fact, in two-dimensional theories with non-derivative interactions, all UV divergences are of the trivial tadpole type. In particular,  the quartic interaction leads to the divergent tadpole shown on the left side of Fig.~\ref{Tadpole}, which we cancel with a mass counterterm, 
\be
-\frac{1}{2}\delta m^2 \phi^2~, \  \ \ \ \ \delta m^2 =  \frac{1}{2} m^2 b^2  \int \frac{d^2 p}{(2\pi)^2}\, G(p)~,
\ee
where $G(p)$ is the propagator. Similarly, the  $\phi^{2n+2}$ interaction in the Lagrangian requires a counterterm, 
\be
-\frac{1}{(2n)!}\delta m^2\, b^{2n-2}\phi^{2n}~.
\ee
Summing all these terms,  the full set of counterterms at one loop is,
\be \label{Lctr}
\mL_{\text{ctr}} =  - \frac{\delta m^2}{b^2} \( \cosh( b \phi) - 1\)~.
\ee
Through a redefinition of the mass,  the renormalized sinh-Gordon Lagrangian is the same as the classical sinh-Gordon Lagrangian. Thus, the quantum sinh-Gordon model maintains all the integrability properties of the classical theory. 

\begin{figure}[t]
\centering
\includegraphics[width=5in]{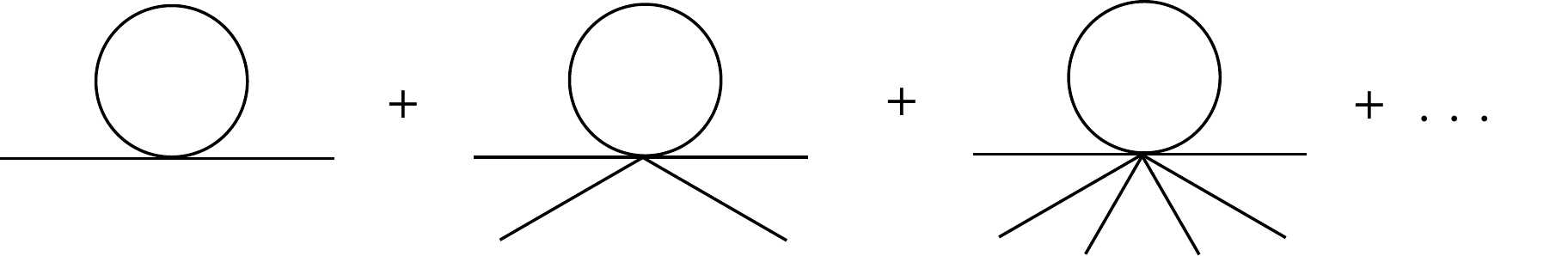}
\caption{The only divergent diagrams in the sinh-Gordon model are  tadpole diagrams. The combinatorics works out so that their only effect is to renormalize the mass. }\label{Tadpole}
\end{figure}
We now turn to the two-to-two  $S$-matrix. For integrable theories, the initial momenta are the same as the final momenta, and so we define the $S$-matrix $S(\theta)$ by, 
\be
{}_{out} \la k_3, k_4| k_1, k_2\ra_{in} = S(\theta)\, (2\pi)^2 2\omega(k_1) 2 \omega(k_2) \delta(k_1 - k_4) \delta(k_2 - k_3)~.
\ee
With this definition, the S-matrix at zeroth order in the coupling is $S^{(0)}(\theta) = 1$. 
The $S$-matrix  is related to the scattering amplitude $\mathcal{A}$ through,
\be \label{AtoS}
S= \frac{\mathcal{A}}{ 4 m^2\sinh\theta}~.
\ee
At tree level, the amplitude is simply  $- i m^2 b^2$, which gives the $S$-matrix, 
\be \label{214}
S^{(1)}(\theta) =  \frac{- i b^2}{4 \sinh \theta}~.
\ee
At  one loop, the $s$-channel amplitude, shown in Fig.~\ref{Oneloop}, is, 
\be
\mA_s = \frac{-m^4 b^4}{2} \int \frac{d^2 q}{(2\pi)^2}\, G(p_1+p_2-q)G(q) =  \frac{i m^4 b^4}{8\pi}\,\frac{ ( i \pi\, - \theta)}{m^2 \sinh \theta}~,
\ee
where the integral is performed in Appendix~\ref{AppendixG}, Eq.~\ref{G12}. Adding also the $t$ and $u$ channels, the one-loop $S$-matrix is thus, 
\be \label{216}
S^{(2)}(\theta) = \frac{b^4}{32\pi \sinh \theta}\( i - \frac{\pi}{\sinh \theta}\)~.
\ee
These order $b^2$ (\ref{214}) and order $b^4$ (\ref{216}) contributions to the $S$-matrix  agree with the Taylor expansion of the exact $S$-matrix (\ref{SGS}). One could in principle proceed to any order.

\begin{figure}
\centering
\includegraphics[width=.85in]{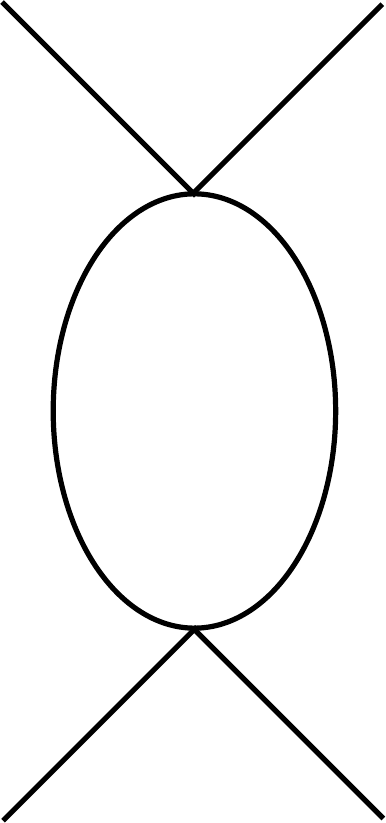}
\caption{The bubble diagram contribution to the sinh-Gordon $S$-matrix. }\label{Oneloop}
\end{figure}

\section{Renormalization of $T\b T$ Deformed Theory} \label{Sec3}
We previously wrote  the classical Lagrangian for the $T\b T$ deformation of a free massive scalar. In this section we  compute the renormalized Lagrangian to order $\lam^2$, at one-loop level. 
The classical Lagrangian was given by Eq.~\ref{Lag}. The first few orders in $\lam$  are,
 \be \label{Lmass}
 \mL =2\d \phi \b \d \phi + \frac{1}{2} m^2 \phi^2 -4\lam (\d \phi \b \d \phi)^2 + \frac{1}{4} \lam m^4 \phi^4 + 16 \lam^2 (\d \phi \b \d \phi)^3 + 2 \lam^2 m^2 (\d \phi \b \d \phi)^2 \phi^2 + \frac{\lam^2}{8} m^6 \phi^6+\ldots~.
 \ee
 The renormalized Lagrangian that we will find is given at the end of the section,  see Eq.~\ref{Lren}. 
 
We will compute the renormalized Lagrangian  by taking the classical Lagrangian, using it to compute the $S$-matrix, and then adding appropriate counterterms so that the $S$-matrix is given by what the definition of the $T\b T$ deformed theories says it should be, Eq.~\ref{STTbar}. Stated differently: as is standard in quantum field theory, in computing the $S$-matrix using the classical Lagrangian, we will encounter UV divergences. We will cancel the divergences through the addition of counterterms. The finite parts of the counterterms, which are usually ambiguous, will be fixed by demanding that integrability be preserved at the quantum level.

For computing the $S$-matrix, we switch to Lorentzian signature, $x_1 = i t$ and $x_2 = x$, so that the holomorphic and antiholomorphic coordinates are, respectively, $z = i \sigma_+$ and $\b z = i \sigma_ - $, where $\sigma_{\pm} = t\pm x$. 
The  action (\ref{Lmass}), to order $\lam^2$,   is thus, 
\bml \label{act}
-I = -\int d x_1 d x_2\,  \mL =
 i \int d t\, dx\, \( 2\d_+ \phi \d_- \phi - \frac{1}{2} m^2 \phi^2 +4\lam (\d_+ \phi  \d_- \phi)^2 -\frac{1}{4} \lam m^4 \phi^4  \right. \\
  \left. + 16 \lam^2 (\d_+ \phi \d_- \phi)^3 - 2\lam^2 m^2 (\d_+ \phi \d_- \phi)^2 \phi^2 - \frac{\lam^2}{8} m^6 \phi^6+ \ldots\)~,
\end{multline}
where we are using the nonstandard convention that  $\d_+$ denotes $\frac{\d }{\d \sigma_+}$. 
The light-cone momenta are $p_{\pm} = \omega \pm k$. In terms of the rapidity $\theta_i$ of particle $i$, the light-cone momenta are $p_{i, \pm} = m e^{\pm \theta_i}$. The Mandelstam variable is $s = p^2$, where $p = p_1+p_2$, is in light-cone variables,  $s = p^2 = p_+ p_-$. Finally, the Feynman propagator is, 
\be
\la \phi(x_1) \phi(x_2)\ra  =  \int \frac{d^2 p}{(2\pi )^2}\, G(p) \, e^{- \frac{i}{2}(p_- \sigma_{12, +} + p_+ \sigma_{12, -}) }~, \ \ \ \ \  G(p)= \frac{i}{p_+p_-   - m^2 + i\eps}~.
\ee

In Sec.~\ref{Massless} we compute the $S$-matrix for the special case of zero mass. In Sec.~\ref{Massive} we consider the massive theory. 

\subsection{Renormalization  of $T\b T$ deformation of massless free scalar} \label{Massless}
In the massless limit, the Lagrangian (\ref{Lag}) becomes \cite{Tateo16}, 
\be \label{NG}
 \mL = \frac{1}{2\lam} \( - 1 + \sqrt{ 1 + 8 \lam\, \d \phi\, \b \d \phi}\)~.
\ee
This is of course the gauge-fixed Nambu-Goto action for a string embedded in three spacetime dimensions, with embedding coordinates $X^{\mu} (\tau, \sigma)$ given by: $X^0 = \tau$,  $X^1 = \sigma$, and $X^2 = \sqrt{2 \lam}\, \phi$.~\footnote{There is an important distinction between this theory and the standard Nambu-Goto theory, defined as the area swept out by a string in $D=3$ embedding spacetime dimensions. For the  Nambu-Goto theory, there is $D$ dimensional Poincare invariance, which must be preserved under quantization.  For instance, in the standard quantization of string theory in light-cone gauge, $D$ can be anything; however, the spectrum is only consistent with Poincare symmetry if $D=26$ (or $D=3$, which is exceptional \cite{Townsend}). In the context of the  $T\b T$ deformation of $D-2$ free scalars, we have no such restriction, as there was no Poincare symmetry to start with. Physically, the action (\ref{NG}) describes small fluctuations around a static long string. The $S$-matrix for Nambu-Goto was   recently studied in \cite{Conkey:2016qju}.}

To study the $S$-matrix in the massless limit, we hold $s$ fixed while taking $m$ to zero and $\theta$ to infinity. 
In this limit, the $S$-matrix (\ref{STTbar}) becomes, 
\be \label{massless}
S = \exp\(i \lam\, s/2\)
~.
\ee
At zero coupling our action is that of a free massless scalar, $\mL= 2 \d \phi \b \d \phi$, 
with correlation functions, $\la \phi(z_1) \phi(z_2)\ra = - \frac{1}{4\pi}\log z_{12}\b z_{12}$.
Since the interactions involve only $\d \phi$ and $\b \d \phi$, it is convenient to think of our fundamental fields as $\d \phi$ and $\b \d \phi$, with propagators given by, 
\be \label{2pt0}
\la \d \phi(z_1)\, \d \phi(z_2)\ra = - \frac{1}{4\pi } \frac{1}{z_{12}^2}~, \ \ \ \ \ \ \la \b \d \phi(z_1)\, \b \d \phi(z_2)\ra = - \frac{1}{4\pi } \frac{1}{\b z_{12}^2}~, \ \ \ \ \la \d \phi(z_1)\, \b \d \phi(z_2)\ra = \frac{1}{4} \delta^2(x_{12})~,  
\ee
where $z_{i j} \equiv z_i - z_j$, and to get the last equation we used that $\b \d \frac{1}{z} = 2\pi \delta^2(z)$.

Let us now compute the $S$-matrix, to order $\lam^2$. The $S$-matrix is related to the amplitude through $S  = \mathcal{A}/2s$. At tree level, the interaction $- 4 \lam (\d\phi \b \d \phi)^2$ in the Lagrangian gives an amplitude $\mathcal{A} = i\lam s^2$. Correspondingly,  the first order $S$-matrix is, 
\be
S^{(1)}= i \frac{\lam}{2} s~,
\ee
in agreement with (\ref{massless}).

\begin{figure}[t]
\centering
\includegraphics[width=.85in]{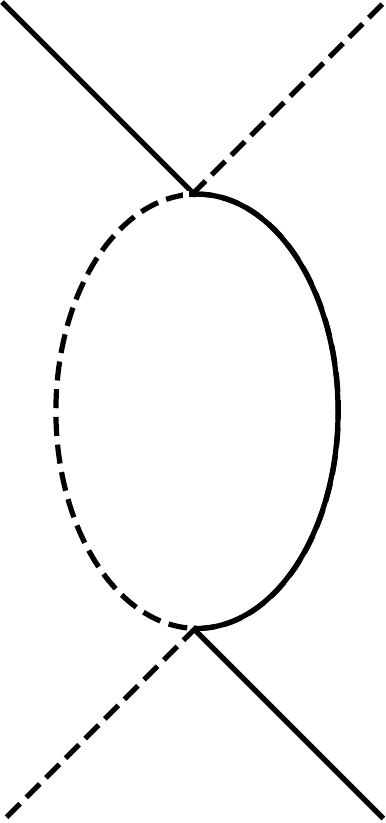}
\caption{The one-loop contribution to the $S$-matrix in the $T\b T$ deformation of a massless free scalar (the gauge-fixed Nambu-Goto theory). Solid lines represent propagators of $\d \phi$, while dashed lines are propagators of $\b \d \phi$. }\label{OneloopMassless}
\end{figure}

At one-loop level, we must compute the bubble diagram shown in Fig.~\ref{OneloopMassless}, where each interaction vertex is $- 4 \lam (\d\phi \b \d \phi)^2$. Since the particles are massless, one of the two incoming particles is  left moving, while the other is  right moving. This uniquely fixes the contractions between the ingoing and outgoing particles and the $\d \phi$ and $\b \d \phi$ that appear in the interaction vertices. Within the loop, along one of the lines we have  a propagator for $\d \phi$, and along the other line a propagator for $\b \d \phi$.~\footnote{The contribution of the mixed propagator $\la \d \phi \b \d \phi\ra$ can be ignored, as it leads to a local and divergent contribution to the amplitude.} Thus, the one loop diagram is, in Euclidean signature, 
\be
L(p_E^2) = \frac{1}{(4\pi)^2}\int d^2x \, e^{i p_E \cdot x}\frac{1}{z^2 \b z^2} =\frac{1}{8\pi}\int \frac{d r}{r^3} \,J_0(|p_E r|)\,~,
\ee
where $p = p_1 + p_2$, and the subscript $E$ denotes Euclidean, and $J_0(r)$ is the Bessel function. We first regulate the integral by putting a hard cutoff, $r=1/\Lambda$, then evaluate the integral, and then take $\Lambda\rightarrow \infty$, to get, 
\be
L(p_E^2) =\frac{1}{8\pi}\int_{\Lambda^{-1}}^{\infty} \frac{d r}{r^3} J_0(|p_E r|)\,  = \frac{\Lambda^2 }{16 \pi}- \frac{p_E^2}{32\pi} (1-\gamma + \log 2) + \frac{ p_E^2}{64\pi} \log\( \frac{p_E^2}{\Lambda^2}\)~.
  \ee
  Converting from Euclidean to Lorentzian signature, using $dt  dx = - i d x_1 d x_2$ and $p_E^2  = -p^2$, the $s$-channel amplitude is thus, 
   \be
\mathcal{A}_s =- 16\, \lam^2\, s^2\,  L(s)~, \ \ \ \ \ \  L(s) =  -\frac{i\, \Lambda^2}{16 \pi} -\frac{i\, s}{32\pi}(1-\gamma + \log 2) +\frac{i\, s}{64\pi}\log\( \frac{-s}{\,\Lambda^2}\)~.
 \ee
Adding the $t$-channel amplitude (where $t=-s$), and dividing by $2 s$ to convert to the $S$-matrix, the second order $S$-matrix element is thus, 
\be
S^{(2)} = i \lam^2 \frac{\Lambda^2}{\pi} s-\frac{\lam^2}{8} s^2~.
\ee
The real piece matches  the order $\lam^2$ part of the $S$-matrix that we want, see Eq.~\ref{massless}.~\footnote{The real piece comes from a $- i\pi$ that comes from the $\log(-s)$ term. Namely, replacing $s$ with $s+ i\eps$, we have $\log(-s- i\eps) = - i \pi + \log(s)$. The $\log(s)$ term gets canceled by the corresponding $t$-channel contribution.} On the other hand, the imaginary and power law divergent piece of $S^{(2)}$ will need to be canceled by a counterterm of the form $(\d \phi \b \d \phi)^2$. This  corresponds to a fairly trivial renormalization of $\lam$.

\subsubsection{Quantum effective action}

In fact, if instead of the on-shell $S$-matrix,  we study the quantum effective action directly, by integrating out high energy modes, we find a number of interesting counterterms. As we will show, these terms do not contribute to the $S$-matrix at second order in $\lambda$. However, they are necessary in the computation of off-shell quantities, such as correlation functions, and play a role in the calculation of higher order terms in the $S$-matrix.

Let us introduce an external field configuration $\b \phi$, through a shift of the field,  $\phi\to \phi + \bar\phi$. The $\lambda^2$ contribution to the effective action is given by the one particle irreducible correlation function of the form, 
\be
 \Gamma_\text{eff}^{(2)}(\bar\phi)=
 8 \lambda^2 \int d^2 x_1 \int d^2x_2\, 
 \big\langle T\bar T(z_1 \bar z_1)  \, T\bar T(z_2,\bar z_2)  \big\rangle_\text{1PI} ~,
\ee
with either $n=0, 2$ or $4$ external legs associated with the background field $\bar\phi$. We denote these terms by $V_n$. Thus, for instance, $V_2$ is given by,
\be
 V_2= -{\lambda^2\over \pi^3} \int d^2 x_1 \int d^2 x_2 \,  \Big[ \del\bar\phi_1 \del\bar\phi_2 {1\over \bar z_{12}^{\, 4}  z_{12}^2} + \text{c.c.} \Big]~,
\ee
where we made use of the correlation functions (\ref{2pt0}), and $\phi_i$ denotes $\phi(z_i)$. This integral exhibits a UV divergence when the points $z_1$ and $z_2$ collide. (In addition, there is a spurious infrared divergence, which is regulated by the decaying external background $\bar\phi$.) The UV divergence is cured in the standard way, through the addition of a local counterterm, $I_\text{c.t.}$,  to the action. In order to obtain the structure of the counterterm, we expand $\bar\phi_2$ around $z_1$,
\be
 V_2= -{\lambda^2\over 3 \pi^3} \int d^2 x_1 \, \del\bar\phi_1 \bar\del^3\del^2 \bar\phi_1   \int d^2 x_2 {1\over  | z_{12}|^2} ~,
\ee
and extract the UV divergent part of the $x_2$ integral. Introducing a sharp UV cutoff $\Lambda$ yields, 
\be
  \int d^2 x_2\, \frac{1}{|z_{12}|^2} = 2\pi \int_{\Lambda^{-1}} {dr\over r}=2\pi \log\Lambda + \ldots~,
  \label{reg_int}
\ee   
where the  ellipsis encode finite contributions. Thus, the divergent part of the two-point vertex takes the form,
\be
 V_2^\text{div}= {2 \lambda^2\over 3 \pi^2} \log\Lambda \int d^2 x ~ \bar\del\del\phi \, (\bar\del\del)^2 \phi  ~,
\ee
where the argument of the logarithm can be made dimensionless by dividing $\Lambda$ by some IR scale. Similarly, the four-point vertex is given by,
\bea \nn
 V_4&=&8\lambda^2 \int d^2x_1 \int d^2 x_2 \, \Big[ \del\bar\phi_1\bar\del\bar\phi_1 \del\bar\phi_2\bar\del\bar\phi_2 {16 \over (4\pi)^2 |z_{12}|^4} + (\del\bar\phi_1)^2(\del\bar\phi_2)^2 {2\over (4\pi)^2 \bar z_{12}^{\, 4}} 
 \\
 && \quad\quad\quad\quad\quad\quad\quad\quad\quad\quad +(\bar\del\bar\phi_1)^2(\bar\del\bar\phi_2)^2 {2\over (4\pi)^2  z_{12}^4} \Big] ~.
\eea
As before, we Taylor expand $\bar\phi_2$ around $z_1$ and integrate over $x_2$ to get the structure of possible UV divergences. Only one divergent integral, which takes the form \reef{reg_int}, survives. In the minimal subtraction scheme, we thus end up with the following set of counterterms, 
\be
 I_\text{c.t.}=V_2^\text{div}+V_4^\text{div} = \lambda^2 \log\Lambda \({2\over 3 \pi^2} \int d^2 x ~ \bar\del\del\phi \, (\bar\del\del)^2 \phi
 +{16\over\pi} \int d^2 x\,  \(\del\phi\bar\del\phi\)  \del\bar\del \(\del\phi\bar\del\phi\) \).
\ee
In fact,  both counterterms are proportional to the equations of motion, $\bar\del\del\phi=0$. For the first term this is manifest, while for the second term  this requires a few manipulations, 
\be
\del \(\del \phi \bar \del \phi\) \bar \del \( \del \phi \bar \del \phi\) \sim \( \del^2\, \bar \del \phi\) \(\del \phi\, {\bar \del}^2 \phi\)\sim \frac{1}{4} \del (\del \phi)^2\, \bar \del (\del \phi)^2 \sim \frac{1}{4} \bar\del (\del \phi)^2\, \del (\del \phi)^2\sim 0~.
\ee
Here in the first equality we acted with the derivatives inside the parentheses and used the equations of motion. We then regrouped the terms to get the second equality, and then integrated by parts in both directions. The equality with zero then follows by the equations of motion, provided the field is on-shell. The effective action is  consistent with our computation of the $S$-matrix, as these terms do not  contribute to the perturbative $S$-matrix at order $\lambda^2$; they vanish when contracted with  on-shell external particles obeying $\del \bar \del \phi=0$.

\subsection{Renormalization of $T\b T$ deformation of massive free scalar} \label{Massive}
In this section we compute the $S$-matrix, to order $\lam^2$, of the $T\b T$ deformation of a free massive scalar. The relevant part of the bare action was given in (\ref{act}). 

 At tree level, there are two contributions to the amplitude. The first is  from the quartic interaction without derivatives, which gives a contribution to the  amplitude of $- 6 i\lam m^4$. The second is from the quartic interaction with derivatives, which gives a contribution to the amplitude of $2 i \lam m^4(2 \cosh^2\theta+ 1)$. Converting from the amplitude to the $S$-matrix via (\ref{AtoS}), the tree level $S$-matrix is trivially seen to be the order $\lam$ piece of (\ref{STTbar}}), 
 \be
 S^{(1)}(\theta) =i \lam m^2 \sinh \theta~.
 \ee
 
At one loop, there are two contributions to the amplitude. The first comes from the tadpole diagram, and the second comes from the bubble diagram.
\subsubsection*{Quartic tadpole diagrams}
\begin{figure}[t]
\centering
\includegraphics[width=1.2in]{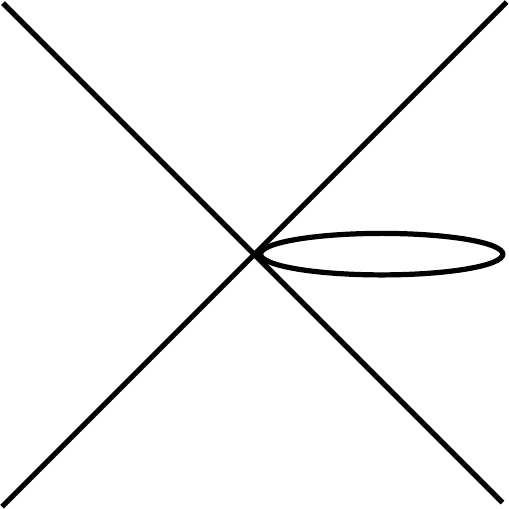}
\caption{Tadpole diagram contribution to the $S$-matrix of the $T\b T$ deformation of a massive free scalar.  }\label{tadMNG}
\end{figure}
From the sextic terms in the Lagrangian, we  have one-loop tadpole diagrams that contribute to the $S$-matrix, see Fig.~\ref{tadMNG}.~\footnote{
There are also tadpole diagrams that renormalize the mass; these are trivial to remove with mass counterterms, and we  simplify notation by taking the mass in the Lagrangian to be the physical mass.}
From the last three terms in the action (\ref{act}), we get the  quartic effective action, 
\be \nn
i \mathcal{L}_{\text{tad}} = \lam^2\(  2 (\d_+ \phi \d_- \phi)^2 \( 72\la \d_+ \phi \d_- \phi\ra - m^2 \la \phi^2\ra  \) - 8 m^2 \d_+ \phi\, \d_- \phi\, \phi^2 \la \d_+ \phi \d_- \phi\ra - \frac{15}{8}  m^6 \phi^4 \la \phi^2\ra \)~.
\ee
The corresponding amplitude is, 
\be
\mathcal{A}_{\text{tad}} = i \lam^2m^4 \Big(\(2 + \cosh 2\theta\) \( 72\la \d_+ \phi \d_- \phi\ra - m^2 \la \phi^2\ra  \)  - 16 \la \d_+ \phi \d_- \phi\ra - 45 m^2 \la \phi^2\ra\Big)~.
\ee
Now, using,
\be
\la \phi^2 \ra = \frac{-1}{4\pi} \log\(\frac{m^2}{\Lam^2}\)~, \ \ \ \  \la \d_+ \phi \d_- \phi\ra= \frac{m^2}{4} \la \phi^2 \ra - \frac{\Lam^2}{16\pi}~,
\ee
where the first equation is found in Appendix~\ref{AppendixG}, Eq.~\ref{C6}, and the second is immediate after writing $p^2 = p^2 - m^2 + m^2$, we get that the tadpole diagram contribution to the amplitude is,   
\be \label{tad}
\mathcal{A}_{\text{tad}}  = \frac{i \lam^2 m^6}{4\pi} \(15 - 17 \cosh 2 \theta\) \log\(\frac{m^2}{\Lam^2}\) - \frac{i \lam^2 m^4 \Lam^2}{2\pi} \(16 + 9 \cosh 2\theta\)~.
\ee

\subsubsection{Bubble diagram}
\begin{figure}[t]
\centering
\subfloat[]{
\includegraphics[width=1.16in]{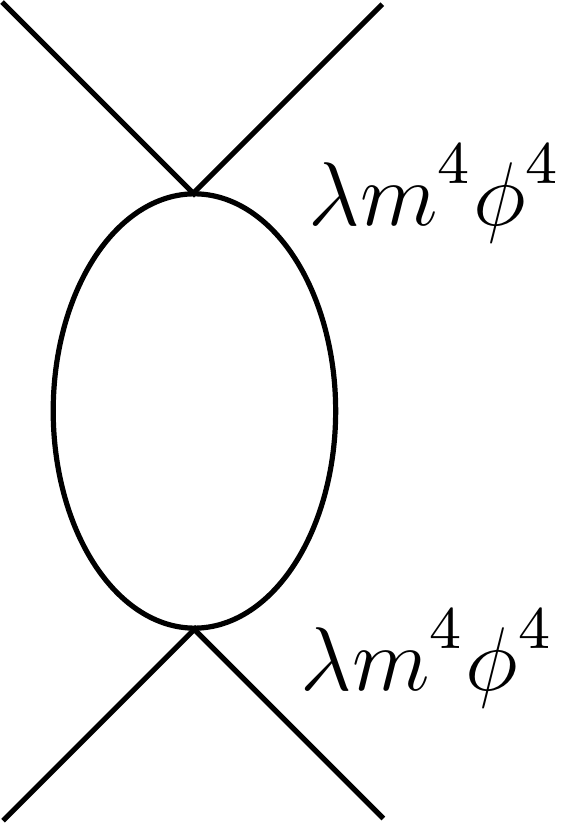}
} \ \ \ \ \ \ \ \ \ \ \ \ \ \ 
\subfloat[]{
\includegraphics[width=1.56in]{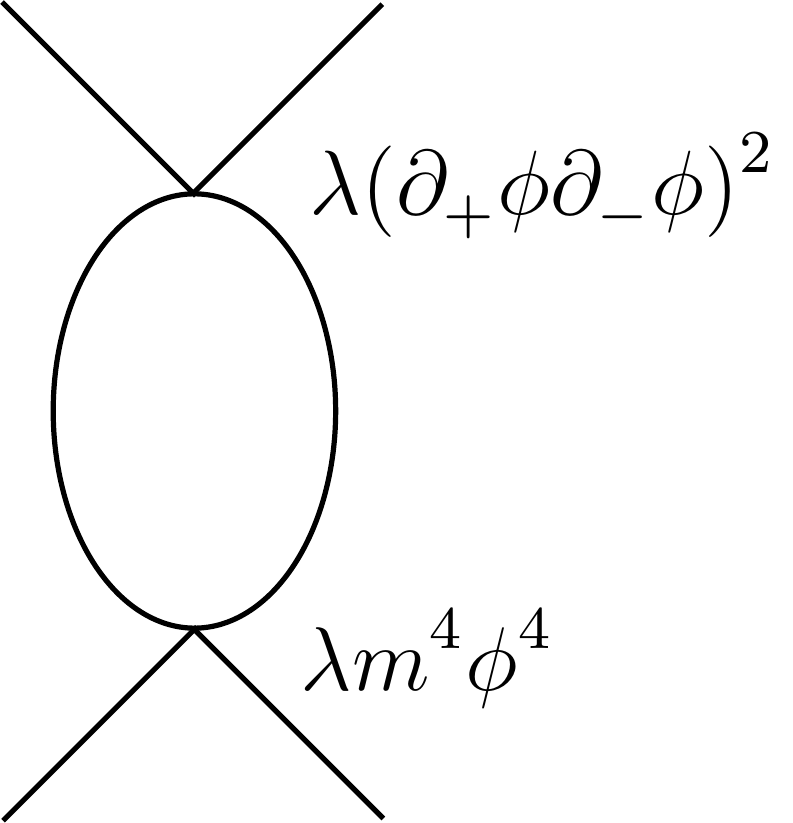}
} \ \ \ \  \
\subfloat[]{
\includegraphics[width=1.56in]{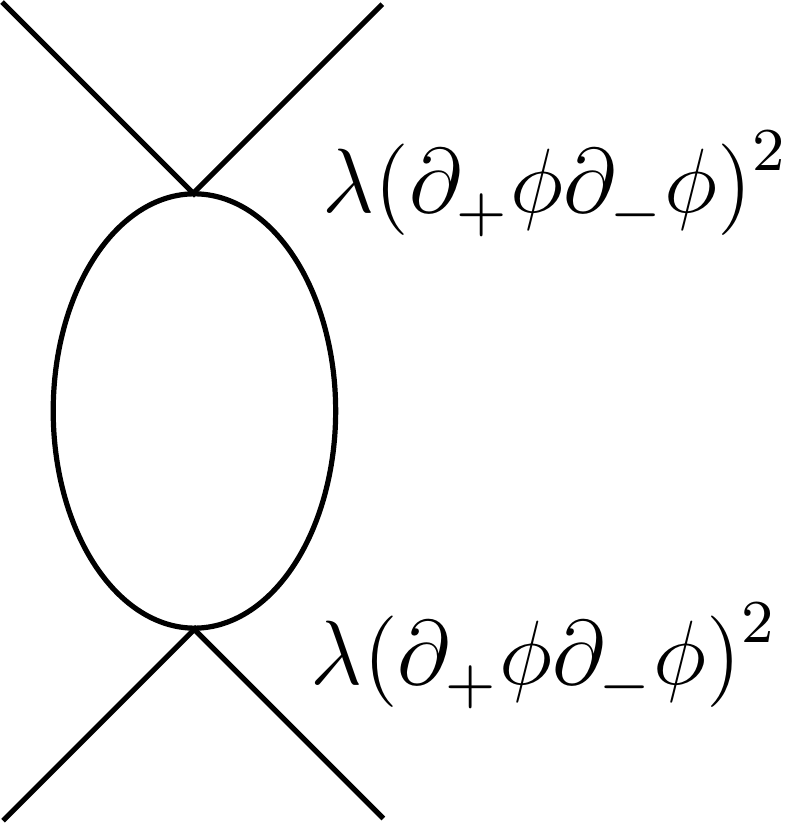} 
}
\caption{The contribution of $s$-channel bubble diagrams to the $S$-matrix of the $T\b T$ deformation of a  massive free scalar. The diagrams are: a) $\mathcal{A}_s^{(a)}$, b) $\mathcal{A}_s^{(b)}$, c) $\mathcal{A}_s^{(c)}$. }\label{Oneloop3}
\end{figure}

We now compute the contribution of the bubble diagram to the $S$-matrix. Since there are two quartic interactions in the Lagrangian (\ref{act}), one with derivatives and one without, we need to consider the bubble diagram in three separate cases, depending on which of the two is chosen for each of the vertices, see Fig.~\ref{Oneloop3}.
\subsubsection*{a) No derivative}
If neither vertex has derivatives, then the amplitude is the same as the corresponding one for the sinh-Gordon model, computed in Sec.~\ref{Sinh}, 
provided that  the coupling $(m^2 b^2/4!)$ is replaced by $\frac{1}{4} \lam m^4$. This gives,
\be
\mathcal{A}^{(a)} =  \frac{9}{2 \pi} \lam^2 m^6 \( i - \frac{\pi}{\sinh \theta}\)~.
\ee

\subsubsection*{b) One derivative}
Next, we look at the bubble diagram with one vertex having derivatives. We get for the $s$-channel amplitude, 
\be
\mathcal{A}^{(b)}_s = - 48 \lam^2 m^6 \( \frac{1}{2} e^{- (\theta_1+ \theta_2)}L_{++}  +\frac{1}{2} e^{ (\theta_1+ \theta_2)} L_{- -} + 2 \cosh\theta  L_{- +} \)~,
\ee
where $L_{\mu \nu}$ is defined by the integral, 
\be
L_{\mu \nu} =\frac{1}{4} \int \frac{d^2 k}{(2\pi)^2} \frac{k_{\mu} }{k^2 - m^2} \frac{(p_{\nu} - k_{\nu})}{(p-k)^2 - m^2}~.
\ee
The integral $L_{ \mu \nu}$ is computed in Appendix~\ref{AppendixG}, see  (\ref{L++v2}). Combined with the trivial relations, $
p_+^2 e^{- (\theta_1+ \theta_2)}= s$ and  $p_-^2 e^{\theta_1 + \theta_2} = s$.
we get for the amplitude,
\be
\mathcal{A}^{(b)}_s =  \frac{3 i \lam^2 m^6}{\pi} \( 1-(2 + \cosh 2\theta)\frac{(i \pi- \theta)}{\sinh \theta} +2  \cosh\theta \log \frac{m^2}{\Lam^2}\)~.
\ee
Adding to this the amplitude in the $t$ channel, $\theta\rightarrow i \pi -\theta$ (note that $\cosh(i \pi - \theta) = - \cosh \theta$ and $\sinh(i \pi - \theta) = \sinh \theta$),  and the amplitude in the $u$ channel, $\theta = i \pi$, we get  the  amplitude, 
\be
\mathcal{A}^{(b)} = \mathcal{A}^{(b)}_s + \mathcal{A}^{(b)}_t + \mathcal{A}^{(b)}_u = - \frac{3 i \lam^2 m^6}{\pi}\( 2 \log \frac{m^2}{\Lam^2} + i \pi \(2 \sinh \theta + \frac{3}{\sinh \theta}\)\)~.
\ee

\subsubsection*{c) Two derivatives}
Finally, we look at the bubble diagram with both vertices having  derivatives. We get for the $s$-channel amplitude, 
\be 
\mathcal{A}^{(c)}_s = 16 \lam^2 m^4\[ \frac{s^2}{p_{+}^4} L_{+ + + +} + 8 \cosh \theta \frac{s}{p_+^2} L_{- + + + } + (1 + 4 \cosh^2 \theta)L_{+ -  +  - } +4 \cosh^2 \theta\, L_{+ + - -} \]~,
\ee
where we defined the integral,
\be
L_{\mu \nu \al \beta}=\frac{1}{16} \int \frac{d^2 k}{(2\pi)^2} \frac{k_{\mu} k_{\nu} }{k^2 - m^2} \frac{(p_{\al} - k_{\al}) (p_{\beta} - k_{\beta})}{(p-k)^2 - m^2}~.
\ee
The integral  $L_{\mu \nu \al \beta}$ is computed in Appendix~\ref{AppendixG}, see  (\ref{L++++}). Using this to get $\mathcal{A}^{(c)}_s$, and then adding the $s$, $t$, and $u$ channels, $\mathcal{A}^{(c)} = \mathcal{A}^{(c)}_s + \mathcal{A}^{(c)}_t + \mathcal{A}^{(c)}_u$,  gives, 
\bml
\mathcal{A}^{(c)} = \frac{i\lambda^2 m^6}{12 \pi } \left(   \pi i \frac{6(2+ \cosh 2\theta)^2}{\sinh \theta}  + 8 + 57 \frac{\Lam^2}{m^2}+ 90 \log \frac{m^2}{\Lam^2}  + 8 \cosh 2\theta\( -2 + 3\frac{\Lam^2}{m^2} + 3 \log \frac{m^2}{\Lam^2} \)\right)~.
\end{multline}

\subsubsection*{All terms}
The full amplitude coming from  the bubble diagram is the sum of these three contributions, $\mathcal{A}_{\text{bub}}= \mathcal{A}^{(a)} +  \mathcal{A}^{(b)}+ \mathcal{A}^{(c)}$,
and is, 
\be\label{bub}
\mathcal{A}_{\text{bub}} = -2 \lam^2 m^6 \sinh^3 \theta + \frac{i \lam^2 m^6}{12\pi} \( 62 + 57 \frac{\Lam^2}{m^2} + 18 \log \frac{m^2}{\Lam^2} + 24 \cosh 2\theta\( - \frac{2}{3} + \frac{\Lam^2}{m^2} + \log \frac{m^2}{\Lam^2}\) \)~.
\ee
\subsubsection{Renormalized Lagrangian}
The total one-loop amplitude is the sum of the contribution  $\mathcal{A}_{\text{tad}}$ (\ref{tad}) of the tadpole diagrams and the contribution $\mathcal{A}_{\text{bub}}$ (\ref{bub}) of the bubble diagrams. Adding these two together, and converting to an $S$-matrix, we find that the second order $S$-matrix is, 
\be\nn
S^{(2)}(\theta) = - \frac{1}{2} \lam^2 m^4 \sinh^2 \theta - \frac{i m^4 \lam^2}{48 \pi \sinh \theta}\( -46 + 69 \frac{\Lam^2}{m^2} - 36  \log \frac{m^2}{\Lam^2}+(32 + 60 \frac{\Lam^2}{m^2}+54 \log \frac{m^2}{\Lam^2} ) \sinh^2 \theta\)~.
\ee
In fact, as  discussed in Appendix~\ref{AppendixG}, when evaluating divergent integrals, we discarded finite pieces, so more precisely the order $\lam^2$ piece of the $S$-matrix is, 
\be \label{S2nd}
S^{(2)}(\theta) = - \frac{1}{2} \lam^2 m^4 \sinh^2 \theta - \frac{i m^4 \lam^2}{16 \pi \sinh \theta}\( 23 \frac{\Lam^2}{m^2} - 12 \log \frac{m^2}{\Lam^2}+( 20 \frac{\Lam^2}{m^2}+18 \log \frac{m^2}{\Lam^2} ) \sinh^2 \theta\)~,
\ee
up to finite terms that are of the functional form of those that can be obtained by changing the cutoff $\Lam$ in this expression by a finite amount.

Since the linear term in $\lam$ in the $S$-matrix  (\ref{STTbar}) is purely imaginary, unitarity completely fixes the real part of the $\lam^2$ term in the $S$-matrix: it was guaranteed that the real part of $S^{(2)}(\theta)$ would come out to $-\frac{1}{2} \lam^2 m^4 \sinh^2\theta$. On the other hand, unitarity tells us nothing about the imaginary part of  $S^{(2)}(\theta)$. 

The second order $S$-matrix that we wanted to get is not (\ref{S2nd}), but rather the order $\lam^2$ piece of (\ref{STTbar}), which contains the same real part but a vanishing imaginary part. We obtained (\ref{S2nd}) as the $S$-matrix coming from the classical (bare) Lagrangian (\ref{Lmass}). We will now add counterterms to it in order to cancel off the imaginary pieces of (\ref{S2nd}). 
Our ability to add local counterterms to the Lagrangian in order to get the $S$-matrix that we want 
 is not a generic property, and is special to the theory being integrable. In particular, generically amplitudes contain logarithms. However, with local counterterms we can only cancel  polynomials of the Mandelstam variable $s$ (equivalently, powers of $\sinh \theta$) that may appear in the amplitude; we can never cancel a term involving $\theta$ (which involves  a logarithm, when expressed in Mandelstam variables, see Appendix~\ref{AppendixG}, Eq.~\ref{Slog}). 
Indeed, we found that even though individual contributions to the amplitude had terms involving $\theta$, they canceled from the final amplitude. This is as it should be, since the $S$-matrix in an integrable theory can only have poles and not branch cuts. 

Adding the appropriate counterterms to cancel the imaginary part of the $S$-matrix, 
the renormalized Lagrangian is thus, 
\be \label{Lren}
\mL =2\d \phi \b \d \phi + \frac{1}{2} m^2 \phi^2 -4 g (\d \phi \b \d \phi)^2 + \frac{1}{4} h m^4 \phi^4+ \ldots~,
\ee
where the renormalized couplings are,~\footnote{The renormalized couplings also have finite pieces;  we have not written them, but it is straightforward to include them: in the one-loop integrals that appeared in the computation of  the $S$-matrix, one should simply keep the finite pieces instead of discarding them. If one were to use the renormalized Lagrangian to compute correlation functions, it would be important to have the correct finite pieces. Indeed, our ability to uniquely fix the finite pieces of the counterterms, by matching to the desired $S$-matrix, is essential. For future calculations, it will be better to use dimensional regularization, rather than the hard cutoff that we have used.}
\bea
g &=& \lam + \frac{\lam^2 m^2}{8 \pi} \( 10 \frac{\Lam^2}{m^2} + 9  \log \frac{m^2}{\Lam^2} \)~, \\
h &=& \lam + \frac{\lam^2 m^2}{24 \pi} \( 7 \frac{\Lam^2}{m^2} + 39  \log \frac{m^2}{\Lam^2} \)~.
\eea

Already at  one-loop level, we see that the renormalized Lagrangian  is qualitatively different from the classical Lagrangian. Namely, in the classical Lagrangian (\ref{Lmass}) there is only one scale, $\lambda$, and it appears as the coupling of both quartic terms, $\phi^4$ as well as $ (\d \phi \b \d \phi)^2$. In the renormalized Lagrangian, the couplings $g$ and $h$ for these two quartic terms are different. 
It would be interesting to know what happens at higher orders in $\lam$. In any case, it is clear that quantum integrability gives rise to a more intricate renormalized Lagrangian than the classical Lagrangian coming from classical integrability. This is qualitatively different from what occurs in the sinh-Gordon model where, as we saw in Sec.~\ref{Sinh}, the renormalized Lagrangian takes the same functional form as the classical Lagrangian.

\section{ The Renormalized Lagrangian, the $T \b T$ Flow Equation, and the $S$-matrix} \label{Sec4}

In Sec.~\ref{STTbarAp} we will start with the $T\b T$ flow equation and derive the $S$-matrix of the $T\b T$ deformed theory (regardless of if the original theory is integrable or not). In the previous section, we used the $S$-matrix to get the renormalized Lagrangian for the $T\b T$ deformed theory. In Sec.~\ref{Tmunu} we discuss general aspects of using the renormalized Lagrangian to verify that the $T\b T$ flow equation is satisfied. This then brings us back full cycle: from the flow equation to the $S$-matrix to the renormalized Lagrangian and back to the flow equation. 
In Sec.~\ref{T2pt} we discuss how one can use the renormalized Lagrangian to compute correlation functions in the $T\b T$ deformed theory. In Sec.~\ref{Integrability}  we discuss integrable deformations of a free theory that go beyond the $T\b T$ deformation.

\subsection{From the $T\b T$ flow equation to the $S$-matrix} \label{STTbarAp}
In this section we present a derivation for the change in the $S$-matrix of any theory, integrable or not, under a $T \b T$ deformation.~\footnote{We are grateful to J.~Maldacena for suggesting  all the key steps in the derivation, as well as very helpful discussions.}

We consider the $n$ body $S$-matrix, where for notational simplicity we take all particles to be incoming. Let $S(\theta_i)$ be the $S$-matrix of some theory. The $S$-matrix of the $T\b T$ deformed theory will be shown to be,
\be
S(\theta_i) \exp\( i \lam m^2\, \frac{1}{2}\!\sum_{1\leq i< j\leq n} \sinh \theta_{i j}\)~,
\ee
where $\theta_{i j} \equiv \theta_i - \theta_j$, and $\theta_i$ are assigned to the particles sequentially in order from left to right.~\footnote{In the case of two-to-two scattering, taking $\theta_3 = i\pi + \theta_1$ and $\theta_4 = i \pi + \theta_2$ (we added the $i\pi$ to make the particles outgoing),  gives back the phase factor discussed before, see Eq. \ref{STTbar}, given by $\exp\( i \lam m^2 \sinh \theta_{12}\)$. }
More generally, the change in the $S$-matrix under a $T_{s+1} \b T_{s+1}$ deformation will be shown to be,
\be \label{SCDD}
S(\theta_i) \exp\( i \lam m^{2s}\, \frac{1}{2}\sum_{1\leq i< j\leq n} \sinh (s\theta_{i j})\)~.
\ee
The currents, written as a one-form, along with the corresponding charges, are, 
\bea \label{QQ}
\star j &=& T_{s+1} d z + \Theta_{s-1} d {\b z}~, \  \ \ \ \ \ \ \ Q= \int \star j~, \\
\star \t j &=& \b T_{s+1} d \b z + \b \Theta_{s-1} d z~, \  \ \ \ \ \ \ \ \t Q= \int \star\t j~,
\eea
and the currents are conserved, $d \star\! j = d \star\! \t j =0$. In terms of these one-forms, the $T_{s+1} \b T_{s+1}$ deformation can be written in a simple form, as the wedge product, 
\be
\(T_{s+1} \b T_{s+1} - \Theta_{s-1} \b \Theta_{s-1}\)  dz\wedge d\b z = \star j \wedge \star \t j~.
\ee
In order to show that the $T_{s+1} \b T_{s+1}$ deformation, as defined by (\ref{TsTs}), gives rise to the $S$-matrix (\ref{SCDD}), it is necessary to show that the finite $\lam$ theory has the matrix element, 
\be \label{412}
\la \text{out}| \int  \star j \wedge \star \t j \, \, | \text{in}\ra =-\frac{m^{2s}}{8}\(\sum_{1\leq i < j\leq n} \sinh (s\theta_{i j})\)\, \la \text{out}|  \text{in}\ra~, \ \ \ \ \theta_{i j} \equiv \theta_i - \theta_j~.
\ee
As we said, we will take the out state to be empty, and the in state to have $n$ particles, each with rapidity $\theta_i$, so that $
| \text{in} \ra = |\theta_1, \ldots, \theta_n\ra$. 

\begin{figure}[t]
\centering
\includegraphics[width=3.7in]{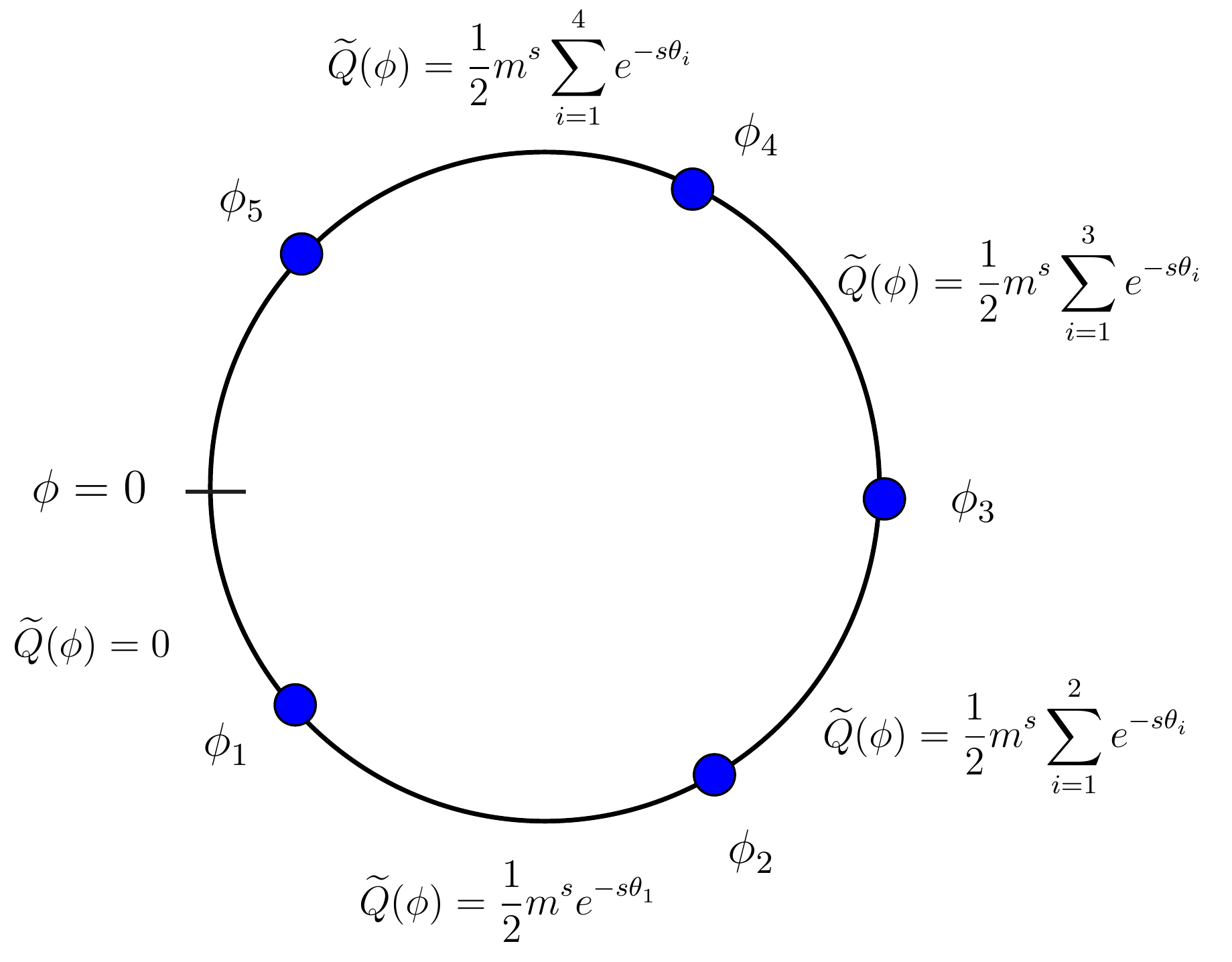}
\caption{The circle is a constant time slice, in radial quantization. We have shown a time slice in the far past (small circle) where the particles are well separated, and represented by the blue dots. A particle at angle $\phi_i$ has rapidity $\theta_i$. $\t Q(\phi)$ is the integral of the antiholomorphic current around the circle, up to angle $\phi$; it undergoes discrete jumps when passing through the particles.}\label{Circle}
\end{figure}
It will be convenient to work in Euclidean signature. It is then natural to use the radial coordinate as the time coordinate: we take the contour of integration  in (\ref{QQ}) to be a circle. We let $\phi$ be the angle around the circle. Furthermore, we define the scalar quantity $\t Q(\phi)$ to be the integral of the current part way around the circle, up to angle $\phi$, 
\be \label{tQd}
\t Q(\phi) = \int_0^{\phi} \star \t j~.
\ee
Notice that $d \t Q(\phi) =  \star \t j$, and in addition, $\t Q(\phi = 2\pi) = \t Q$. Since the current is conserved, we have that $d ( \star j \t Q(\phi)) = \star j\, d \t Q(\phi)$. This allows us to rewrite, 
\be \label{48}
 \int \star j \wedge \star \t j \, = \int  d \( \star j\, \t Q(\phi)\) = \int_{\d}  \star j\,  \t Q(\phi)~.
 \ee
 where in the last step we applied Stokes's theorem. The boundary $\d$ that appears here consists of two disconnected circles: a small circle around the origin (corresponding to a Lorentzian time that is in the far past) and a large circle around the origin (corresponding to a Lorentzian time that is in the far future). Since we have only incoming particles, only the small circle is relevant. 
The left-hand side of (\ref{412}) is thus, 
 \be \label{415}
\la \text{out}| \int  \star j \wedge \star \t j \, \, | \text{in}\ra = \la 0|  \int_{\d}  \star j\,  \t Q(\phi)~ |\theta_1, \ldots, \theta_n\ra~.
\ee
The integral on the right is an integral over the angle $\phi$, around the circle. As we move around the circle, $\star j$ is the current at the angle $\phi$, while $\t Q(\phi)$ defined in (\ref{tQd}) is the integral of the current from $0$ to $\phi$. 
For the moment, suppose we consider a state of a single particle, $|\theta_i\ra$. It is an eigenstate of the charge operator, 
\be \label{410}
Q |\theta_i\ra = \frac{1}{2} m^s e^{s \theta_i} |\theta_i\ra~, \ \ \ \ \t Q |\theta_i\ra =\frac{1}{2} m^s e^{-s \theta_i} |\theta_i\ra~.
\ee
In the case of $s=1$, the charges $Q$ and $\t Q$ are just the holomorphic and antiholomorphic momenta, respectively.
We assume that the state of $n$ particles that we have, $|\theta_1, \ldots, \theta_n\ra$, consists of localized particles at separated angles $\phi_i$, with the particle at $\phi_i$ having rapidity $\theta_i$. This state is an eigenstate of the operator $Q(\phi)$, 
\be
Q(\phi) |\theta_1, \ldots, \theta_n\ra = \frac{1}{2}m^s \sum_{i=1}^k e^{s \theta_i}  |\theta_1, \ldots, \theta_n\ra~, \ \ \ \ \t Q(\phi)|\theta_1, \ldots, \theta_n\ra = \frac{1}{2} m^s\sum_{i=1}^k e^{-s \theta_i}  |\theta_1, \ldots, \theta_n\ra~,
\ee
where $k$ is the maximum index $i$ such that the particle with rapidity $\theta_i$ is at an angle $\phi_i$ for which  $\phi_i< \phi$.  An illustration is shown in Fig.~\ref{Circle}. $Q(\phi)$ is a sequence of step functions, undergoing jumps at $\phi_i$. We assume that the value at precisely $\phi_i$ is given by the average of the value to the left and to the right of $\phi_i$, 
\be
 Q(\phi_i) \equiv \frac{1}{2}\(  Q(\phi_i - \eps) + Q(\phi_i+\eps)\)~. 
 \ee
 When we go fully around the circle and get back to $\phi =0$, we must have that  $0 = Q(0) =Q(2\pi)$, when acting on this state,  which follows from 
charge conservation  (since we assumed that the out state is empty), 
\be \label{cons}
\sum_{i = 1}^n e^{\pm s \theta_i} = 0~.
\ee
We may now evaluate (\ref{415}), 
\bea\nn
 \la 0|  \int_{\d}  \star j\,  \t Q(\phi)~ |\theta_1, \ldots, \theta_n\ra &=& \frac{1}{2} m^s \sum_{i =1}^n e^{s \theta_i} \t Q(\phi_i) = \frac{1}{8}m^{2s }\( \sum_{1\leq j<i\leq n} e^{s \theta_{i j} }+ \sum_{1\leq j\leq i\leq n} e^{s \theta_{i j} }\)\\  \label{415v2}
 & =& -\frac{1}{8} m^{2 s}  \!\!\sum_{1\leq i < j\leq n} \sinh s \theta_{i j}~,
\eea
where in the last step we used that the total charges is zero (\ref{cons}), and thus so is the product of the total left charge and total right charge, $0 =  \sum_{i, j=1}^n e^{s \theta_{i j}}$. We have thus shown (\ref{412}), and hence (\ref{SCDD}).

Although we took all particles to be incoming, the answer is of course valid if some of the particles are outgoing, which one can achieve by sending $\theta$ to $i \pi + \theta$. If one wishes, one can also modify the derivation to make some of the particles  outgoing. Letting the index $a$ denote the ingoing particles and the index $b$ denote the outgoing particles, we get a contribution that is like the one we found before, Eq.~\ref{415v2}, but now with a sum only over the $a$ indices. This came from the small boundary circle. There is also the large boundary circle, as discussed below Eq.~\ref{48}, which gives an additional contribution which is of the form (\ref{415v2}), but with $b$ indices. Thus, we have, 
\be \label{416}
 -\frac{1}{8} m^{2 s} \sum_{a< \t a} \sinh s \theta_{a \t a} -\frac{1}{8} m^{2 s}  \sum_{b< \t b} \sinh s \theta_{b  \t b}~.
 \ee
Multiplying the two momentum conservation equations: $\sum_a e^{\theta_a} = \sum_b e^{\theta_b}$ and $\sum_b e^{-\theta_b} = \sum_a e^{-\theta_a}$ gives that $\sum_{a, b} \sinh \theta_{ab}=0$. Hence, (\ref{416}) reproduces (\ref{415}). 

The $S$-matrix ``dressing'' (\ref{SCDD}), applied to any theory, was first considered in \cite{Dubovsky:2012wk, Dubovsky:2013ira} (for $s=1$). This is naturally the $S$-matrix of a QFT coupled to JT gravity \cite{Dubovsky:2017cnj}. The desired result of this section - that this is also the $S$-matrix of the $T\b T$ deformed theory - then requires showing that the $T\b T$ deformed theory is the same as the theory coupled to JT gravity \cite{Dubovsky:2018bmo}. The derivation in this section is more direct, by bypassing JT gravity, and it is also valid for $T_{s+1} \b T_{s+1}$ deformations.

\subsection{From the renormalized Lagrangian to the  $T\b T$ flow equation} \label{Tmunu}

The $T\b T$ deformed theories are defined by the $T\b T$  flow equation, Eq.~\ref{DiffEI}. From the flow equation one can find the $S$-matrix for the $T\b T$ deformed theories, as we saw in the previous section. Then, from the $S$-matrix, one may hope to construct the renormalized Lagrangian; we did this explicitly in Sec.~\ref{Sec3} for the $T\b T$ deformation of a free scalar. One thing that remains of interest  to verify is that the renormalized Lagrangian satisfies the original $T\b T$ flow equation. In this section, we make some general comments on this.   Everything we will say also trivially extends to the higher-spin $T_{s+1} \b T_{s+1}$  deformations described by (\ref{TsTs}). 

In our setup, we started with the $T\b T$ flow equation interpreted as a classical equation. The solution of this gave the classical Lagrangian, which was our starting point for constructing the renormalized Lagrangian. It is then no longer obvious  how the $T\b T$ flow equation manages to hold as a quantum equation,  with the renormalized Lagrangian on the left and renormalized composite operators on the right. 

First we note that, as a result of the Ward identity, if one has a conserved current, then the renormalized current can  differ from the canonical unrenormalized current only by terms that are themselves conserved.
 For instance, in the context of $\phi^4$ theory in four dimensions, the renormalized stress tensor is, see e.g., \cite{Brown:1979pq, Brown}, 
\be \label{Tphi4}
T^{\mu \nu} = T_0^{\mu \nu} + \eta_0 \(\eta^{\mu \nu} \d^2 - \d^{\mu} \d^{\nu}\) \phi_0^2~,
\ee
where $\phi_0$ is the bare field and $T_0^{\mu \nu}$ is the classical stress tensor, 
\be
T_0^{\mu \nu}  = \d^{\mu} \phi_0 \d^{\nu} \phi_0 - \eta^{\mu \nu} \( \frac{1}{2}(\d \phi_0)^2 + \frac{1}{2} m_0^2 \phi_0^2 + \frac{1}{4!} \lam_0 \phi_0^4\)~.
\ee
Notice that the renormalized stress tensor may differ from the classical stress tensor: one must choose a specific $\eta_0$ so that $T^{\mu \nu}$ is finite. The improvement term on the right-hand side of (\ref{Tphi4}) is itself a conserved quantity, which is why it appears. There are many other  improvement terms, however these all contain higher derivatives, and as our couplings $\lam$ and $m$ are of non-negative dimension, they are excluded on dimensional grounds.

Let us now turn to field theory in two dimensions. In two dimensions,  the equation for conservation of the stress tensor is,
\be
\b \d T = \d \Theta~.
\ee
We may deform the stress tensor by any quantity that is itself conserved. In particular, we may send $T \rightarrow T - \d \rho$ and $\Theta \rightarrow \Theta - \b \d \rho$. Since we need $\Theta = \b \Theta$, we must have $\b \d \rho = \d \b \rho$. Provided that this is satisfied, we can choose any $\rho$, and thereby get a new stress tensor. The renormalized stress tensor must thus take the form, 
\be
T = T_0 -\d \rho~, \ \ \ \ \Theta =   \Theta_0- \b \d \rho~,
\ee
where $T_0$ and $\Theta_0$ are the components of the bare stress tensor, and $\rho$ is something that would  need to be computed for the specific theory.~\footnote{An important distinction with the  $ \phi^4$ theory in four dimensions is that, in the $T\b T$ deformed theories,  the renormalized Lagrangian will acquire new counterterms that were not present in the classical Lagrangian. So, for the purposes of this paragraph, by ``bare stress tensor'' we mean a stress tensor computed classically from a Lagrangian, but the Lagrangian is not the classical Lagrangian but rather the classical Lagrangian plus all terms that will be picked up after renormalization.} Since in the $T\b T$ deformed theory the coupling $\lam$ has negative dimension, there are an infinite number of things that $\rho$ can potentially be. 

In any case, the quantity that we are interested in is the one that appears on the right-hand side of the $T\b T$ flow equation, 
\be
X= T\b T - \Theta \b \Theta~.
\ee
The relation between the bare $X_0$ and the renormalized $X$ is thus, 
\be
X = (T_0- \d \rho)(\b T_0 - \b \d \b \rho) - (\Theta_0- \b \d \rho)^2 = X_0 + \d(\b Y + \rho \b \d \b \rho)+\b \d (Y - \rho \d \b \rho)~,
\ee
where $Y = \rho \Theta_0- \b \rho\, T_0$, and we made use of conservation of the bare stress tensor. Thus, $X_0$ and $X$ differ by a total derivative, which does not impact the action.

\subsection{From the renormalized Lagrangian to correlation functions} \label{T2pt}

Suppose that we wanted to understand the $\lam$ dependence of correlation functions of $\phi$ for the $T\b T$ deformed theory \cite{Kraus,Ofer, Guica:2019vnb,  Cardy:2019qao}. 
A straightforward way of proceeding would be to compute the correlators perturbatively in $\lam$. 
If we were to do this using the bare Lagrangian, then we would expect to find UV divergences. We could cancel these by adding counterterms to the Lagrangian; however, since the finite part of the counterterms is ambiguous, it would naively seem that the correlation functions are ambiguous. It is here that the results of Sec.~\ref{Sec3} become important: we fully fixed, to order $\lam^2$, the renormalized action, by demanding that the corresponding $S$-matrix be $\exp(i \lam m^2 \sinh \theta)$. Assuming that we can uniquely fix the renormalized action at arbitrary order in $\lam$, this then gives a way to obtain unique correlation functions. If, instead of correlation functions of $\phi$, we are interested in correlation functions of composite operators, then we must, in addition, renormalize the composite operators.

\subsubsection{Stress tensor two-point function}
In this section we  evaluate the two-point function of the  stress tensor of the $T\b T$ deformation of a massless scalar, to second order in the coupling $\lam$. This is not a sufficiently high order in $\lam$ for anything interesting to happen since,  as discussed in Sec.~\ref{Massless}, at second order in $\lam$ there are no $\log$ divergences. Thus,  we expect to be able to get a finite answer by simply evaluating the integrals via analytic continuation, with no need for the addition of counterterms. However, the computation we perform is a necessary warmup for the computation at higher orders in $\lam$.

The free massless scalar has a two-point function $\la \phi(z_1) \phi(z_2)\ra = - \frac{1}{4\pi}\log z\b z$ (see Eq.~\ref{2pt0}). The stress tensor $T_{\mu \nu}$ has components $T_{z z} \equiv T$,\, $T_{\b z \b z} \equiv \b T$ and $T_{z \b z}$. For the free scalar, the only nonzero components are $T = (\d \phi)^2$ and $\b T = (\b \d \phi)^2$. Trivially,  the two point function is,
\be \label{T0}
\la T(z_1) T(z_2) \ra = \frac{1}{8 \pi^2} \frac{1}{z_{12}^4}~.
 \ee
For the Nambu-Goto theory, from the action (\ref{NG}), one finds that the stress tensor is, 
\be \label{T}
T^{\lam}= \frac{(\d \phi)^2}{\sqrt{1 + 8\lam \d \phi \b \d\phi}}~, \ \ \ \ \b T^{\lam} = \frac{(\b \d \phi)^2}{\sqrt{1 + 8\lam \d \phi \b \d\phi}}~, \ \ \ \ \ \ T_{z \b z}^{\lam} = \frac{\d \phi \b \d \phi} {\sqrt{ 1 + 8 \lam \d \phi \b \d \phi}} - \frac{1}{2} \mL~.
\ee
In computing the two-point function of the stress tensor, $\la T^{\lam}_{\mu \nu} (z_1) T^{\lam}_{\al \beta}(z_2)\ra_{\lam} $, perturbatively in $\lam$, we must account for  the $\lam$ dependence of $T_{\mu \nu}^{\lam}$, which we expand in $\lam$, $T_{\mu \nu}^{\lam} = T_{\mu \nu} + \lam T^{(1)} + \ldots$, as well as the $\lam$ dependence of the action (\ref{NG}), which we also expand in $\lam$, $I = I_0 + \lam I_1 + \ldots$. Explicitly, we have, 
\bea\nn
T^{\lam} &=& (\d \phi)^2\( 1- 4\lam \d \phi \b \d \phi  + \ldots\)~, \ \  \  {\b T}^{\lam} = (\b \d \phi)^2\( 1- 4\lam \d \phi \b \d \phi  + \ldots\)~,\\
 T_{z \b z}^{\lam}&=&-2 \lam (\d \phi \b \d \phi)^2+ \ldots ~, \ \ \ \ \  \ \ \ \ \ \ \ \ 
I = \int d^2 x \( 2\, \d \phi \b \d \phi - 4\lam (\d \phi \b \d \phi )^2 + \ldots\)~.
\eea
We now proceed to compute, perturbatively in $\lam$, the two-point function  of the stress tensor.  At zeroth order in $\lam$, the result is that of the free scalar (\ref{T0}), while at first order in $\lam$, all contributions manifestly vanish. Thus, we need to look at the order $\lam^2$ contributions. 
We start with the simplest components,  $\la T_{z \b z}^{\lam}(z_1) T_{z \b z}^{\lam}(z_2)\ra $. Since $T_{z \b z}^{\lam}$ starts at order $\lam$, we simply Wick contract to get, 
\be
\la T_{z \b z}^{\lam} (z_1) T_{z \b z}^{\lam}(z_2) \ra_{\lam} =  \frac{\lam^2}{16 \pi^4} \frac{1}{z_{12}^4 \b z_{12}^{\, 4}}~, 
\ee
at order $\lam^2$. This result was found in \cite{Kraus}, and all other components were then  obtained from it through use of energy-momentum conservation. However, it is instructive to calculate these directly as well. 
Proceeding,  we  next look at $\la T^{\lam}(z_1) \b T^{\lam}(z_2)\ra_{\lam}$. The only nonzero contribution is, 
\be 
\!\!\!\!\!\!  \la T^{\lam}(z_1) \b T^{\lam}(z_2)\ra_{\lam} =\! \frac{\lam^2}{2} \la T(z_1) \b T(z_2) I_1 I_1\ra 
= \! \frac{8 \lam^2}{(2\pi)^6} \int\! d^2 x_3 d^2 x_4 \frac{1}{z_{13}^2 z_{34}^2 z_{41}^2}\frac{1}{ \b z_{23}^{\, 2} \b z_{34}^{\, 2} \b z_{42}^{\, 2}} =\!  \frac{\lam^2}{16\pi^4} \frac{1}{z_{12}^4 \b z_{12}^{\, 4}}~,
\ee
where we used (\ref{DI}) to evaluate the double integral. 
Next, we look at $\la T^{\lam}(z_1) T^{\lam}_{z \b z}(z_2)\ra$. The only contribution is, 
\be
\la T^{\lam}(z_1) T^{\lam}_{z \b z}(z_2)\ra_{\lam} = 2\lam \la (\d \phi_1)^2  (\d \phi_2 \b \d \phi_2)^2\, I_1\ra  = \frac{\lam^2}{8 \pi^5} \frac{1}{z_{12}^2} \int d^2 x_3 \frac{1}{z_{13}^2 z_{23}^2 \b z_{23}^{\, 4}} =  -\frac{\lam^2}{12 \pi^4} \frac{1}{z_{12}^5 \b z_{12}^{\, 3}}~,
\ee
where we used (\ref{A4}) to evaluate the integral. 

Finally,  we look at $\la T^{\lam}(z_1) T^{\lam}(z_2)\ra_{\lam}$. There are three distinct contributions. The first comes from the order $\lam$ contribution of $T^{\lam}$,  
\be \label{tree1}
 \la T^{(1)}(z_1) T^{(1)}(z_2)\ra = \frac{6\lam^2}{(2\pi)^4} \frac{1}{z_{12}^6\,  \b z_{12}^{\, 2}}~.
 \ee
 A second contribution comes form the mixing of the order $\lam$ part of the stress tensor and the first order in $\lam$ part of the action,
  \be \nn
- \la T^{(1)} (z_1) T^{(0)}(z_2) I_1\ra - (1\leftrightarrow 2)= -32\lam^2\! \int d^2 x_3 \la (\d \phi_1)^3\,  \b \d \phi_1   (\d \phi_2)^2 (\d \phi_3)^2 (\b \d \phi_3)^2\ra  = \frac{-12 \lam^2}{(2\pi)^4} \frac{1}{z_{12}^6\,  \b z_{12}^{\, 2}}~.
\ee
Here, the only piece that contributes comes from a contact-like term: a contraction  $\la \d \phi_2 \b \d \phi_3\ra$, which is proportional to a delta function (see Eq.~\ref{2pt0}). A third contribution comes from two  first order in $\lam$ pieces of the action, 
 \be \nn
 \la T(z_1) T (z_2) I_1I_1\ra = \frac{4 \lam^2}{ (2\pi)^6} \int d^2 x_3 d^2 x_4\,  \frac{1}{{\b z}_{34}^{\, 4}}\( \frac{1}{(z_{13} z_{14} z_{23} z_{24})^2}+ \frac{2}{(z_{12} z_{14}z_{23} z_{34})^2}  \)+ \frac{6 \lam^2}{(2\pi)^4} \frac{1}{z_{12}^6\,  \b z_{12}^{\, 2}}~,
  \ee
 where the second term comes from the contact-like contractions between $\d \phi$ and $\b \d \phi$. 
 The integrals in the first term are evaluated using the integrals in Appendix~\ref{IntEx}, giving, 
\be
\int d^2 x_3 d^2 x_4\,  \frac{1}{{\b z}_{34}^{\, 4}} \frac{1}{(z_{13} z_{14} z_{23} z_{24})^2} = \frac{4 \pi^2}{3}\frac{1}{z_{12}^6 \b z_{12}^{\, 2}}~, \ \ \ \ \ \int d^2 x_3 d^2 x_4\frac{1}{{\b z}_{34}^{\,4}} \frac{1}{(z_{14}z_{23} z_{34})^2} =  \frac{\pi^2}{z_{12}^4 \b z_{12}^{\, 2}}~.
\ee
Combining all three contributions, in total we have, 
\be
\la T^{\lam}(z_1) T^{\lam}(z_2) \ra_{\lam} = \frac{1}{z_{12}^4}\(  \frac{1}{8 \pi^2}  + \frac{5 \lam^2}{24 \pi^4} \frac{1}{z_{12}^2\b z_{12}^{\, 2}}\)~.
\ee

This completes the computation of the two-point function of $T_{\mu \nu}$ for the  $T\b T$ deformation of a massless scalar.
It would be interesting to repeat the computation for the $T\b T$ deformation of a massive scalar. We save this for future work; although we have the renormalized Lagrangian, a necessary ingredient  in the computation is, in addition, the renormalized $T_{\mu \nu}$ (for the massless case $T_{\mu \nu}$ does not get renormalized at second order).

\subsection{Further integrable deformations of a free theory} \label{Integrability}
In Sec.~\ref{sub1} we show that there is a broad class of integrable deformations of a free theory. In Sec.~\ref{sub2} we argue that this can be thought of as a consequence of the higher spin $T_{s+1} \b T_{s+1}$ deformations, of which $T \b T$ is a subcase. In Sec.~\ref{sub3} we contrast this with theories with nonderivative interactions, like the sinh-Gordon model, which is the unique integrable theory in its class. 

\subsubsection{$\mathcal{L} = f( \lam  \d \phi \b \d \phi)$ is integrable} \label{sub1}
An important property of  $T\b T$ is that the $T \b T$ deformation of an integrable theory is also integrable.  An example that we studied in Sec.~\ref{Massless} is the $T \b T$ deformation of a free massless scalar, corresponding to a classical action that is the gauge fixed Nambu-Goto action (\ref{NG}). In this section we  show that, in fact, any action of the  general form, 
\be \label{426}
\mathcal{L} = f(x)~, \ \ \ \ x = \lam\, \d \phi \b \d \phi~,
\ee
where $f(x)$ is any function that is analytic in the vicinity of $x=0$, is classically integrable.

To show that the theory is integrable,  we need to find an infinite number of conserved currents, $(T_n, \Theta_{n-2})$, which we parametrize as, 
\be
T_n = (\d \phi)^n\, t_n(x)~, \ \ \ \ \ \ \Theta_{n-2} = (\d \phi)^{n-1} (\b \d \phi)\, h_{n }(x)~.
\ee
The conservation equation  $\b \d T_n = \d \Theta_{n-2}$  becomes, 
\be \label{59}
\frac{x}{\lam} \d \b \d \phi \[n t_n  - h_n + x t_n' - x h_n'\] = - (\d \phi)^2 {\b \d}^2 \phi\, x t_n' + (\b \d \phi)^2 { \d}^2 \phi\( (n-1) h_n + x h_n'\)~.
\ee
We need this to be consistent with the equations of motion, which are given by, 
\be \label{54}
\d \b \d \phi = - \frac{\lam}{2} \frac{f''(x)}{f'(x) + x f''(x)}\( (\d \phi)^2\,  \b \d^{\, 2} \phi + (\b \d \phi)^2\, \d^2 \phi\)~.
\ee
Thus, we need to have that, 
\be \label{510}
- x t_n' =  (n-1) h_n + x h_n'~.
\ee
Inserting this into (\ref{59}), we see that consistency with the equations of motion further requires,
\be \label{513}
n t_n  - h_n + x t_n' - x h_n' =2\frac{f'(x) + x f''(x)}{f''(x)}~.
\ee
We can turn these two equations, (\ref{510}) and (\ref{513}),  into one equation. Solving (\ref{510}) for $t_n$ gives, 
\be
t_n = - h_n + (n-1) \int^x dy \, \frac{h_n(y)}{y}~. 
\ee
Inserting this into (\ref{513}) gives an integral equation for $h_n$, 
\be
-  x h_n' -  n h_n + \frac{1}{2}n(n-1) \int^x dy\, \frac{h_n(y)}{y} = \frac{f'(x) + x f''(x)}{f''(x)}~.
\ee
Given a Lagrangian $\mathcal{L} = f(x)$,  this can be explicitly solved to obtain $h_n$. Therefore, the theory is classically integrable. 

\subsubsection{$T_{s+1} \b T_{s+1}$ deformations of a free massless scalar} \label{sub2}
In addition to the $T\b T$ deformation, there are also the  higher spin $T_{s+1} \b T_{s+1}$ deformations, defined earlier in (\ref{TsTs}).~\footnote{Other variants of $T\b T$ include supersymmetric versions \cite{Baggio:2018rpv,Chang:2018dge},  the $J \b T$ deformation \cite{Guica:2017lia}, simultaneous $J \b T$ and $T\b T$ deformations \cite{Mark, Frolov:2019nrr},  $T\b T$ in curved space \cite{Jiang:2019tcq}, and deformations in quantum mechanics \cite{David}. See also \cite{LeFloch:2019wlf, Sfondrini:2019smd}.} These are like the $T\b T$ deformations, but with the higher spin currents, rather than the stress tensor. They too preserve integrability. One may want to view  the general integrable theory that we just discussed, $\mathcal{L} = f(x)$, as being formed from a superposition of $T_{s+1} \b T_{s+1}$ deformations of a free massless scalar. 

The classical Lagrangian for the $T_{s+1} \b T_{s+1}$ deformation of a free massless scalar is straightforward to find: one solves the $T_{s+1} \b T_{s+1}$ flow equation (\ref{TsTs}) as well as the conservation equation  $\b \d T_n = \d \Theta_{n-2}$, perturbatively in $\lam$. The resulting Lagrangian is \cite{KR},
\be
\mL_s= \sum_{n=0}^{\infty}  c_{n} \lam^n (\d \phi \b \d \phi)^{ns+1}~,
\ee
where the coefficients $c_n$ are given recursively through, 
 \bea
 c_n &=&  \sum_{m=0}^{n-1} \frac{2^{n+1}}{(m+1)(m-n)}\, d_m d_{n-1-m}~,\\
  d_p   &=&\frac{1}{ 2 p } \sum_{m=1}^p c_m\, d_{p-m}\, \frac{(1+m s)( 4 m^2 + 2 p (1+p) - m (3+ 6 p + s))}{m-p-1}~.
  \eea
Explicitly, the first several coefficients are, 
  \be
c_{0} = 2,~ \  \  \ c_{1} = -4~, \ \ \ \ c_{2} =4 (s+1)^2 ~, \ \ \ \ c_{3} =  -(7s+3)(s+1)^3~.
\ee
For the special case of $s=1$ (the $T\b T$ deformation), this reproduces the Nambu-Goto Lagrangian (\ref{NG}) found in \cite{Tateo16}. For general $s$, there does not appear to be a simple closed-form expression for the Lagrangian.

\subsubsection{Uniqueness of sinh-Gordon} \label{sub3}
One of the motivations for this section was the well-known property of uniqueness of the sinh-Gordon model. More precisely, within the class of scalar field theories, 
\be \label{438}
\mL = 2 \d \phi \b \d \phi + V(\phi)~,
\ee
with symmetry $\phi \rightarrow - \phi$, the sinh-Gordon model is the unique integrable theory. A simple argument for this is to look at the constraints that the existence of a higher spin conserved current imposes on $V(\phi)$ \cite{Bajnok}.~\footnote{An alternative is to consider the Taylor expansion of $V(\phi)$ and fix the coefficients by demanding that the tree level $2$ to $n$ $S$-matrix be zero for all $n>2$ \cite{Arefeva:1974bk, Dorey:1996gd}.  As a practical matter, this is significantly less efficient; see however \cite{Gabai:2018tmm}.} 

The equations of motion are $\d \b \d \phi = V'$. Correspondingly, the energy-momentum tensor is conserved, 
\be
\b \d T_2 = \d \Theta_0~, \ \ \ T_2 = \frac{1}{2} (\d \phi)^2~, \ \ \ \Theta_0 = V~.
\ee
Next, looking at the spin $4$ current, it must take the form,
\be
T_4 = \frac{1}{4} (\d \phi)^4 + \alpha (\d^2 \phi)^2~,
\ee
where $\alpha$ is some parameter. We get   $\Theta_2$ by imposing conservation of the current, $\b \d T_4 = \d \Theta_2$, which yields, 
\be
\b \d T_4 = \d \( (\d \phi)^2 V\) +2 (\d^2 \phi ) (\d \phi)\( \alpha V'' - V\)~.
\ee
For this to take the required form, $\b \d T_4 = \d \Theta_2$, 
we must have $\( \alpha V'' - V\) = 0$; the solution of which is the sinh-Gordon potential. 

The reason integrability for Lagrangians of the form $\mathcal{L} = f( \lam  \d \phi \b \d \phi)$ is so much less constraining than for Lagrangians of the form (\ref{438}) is clear: for the former, unlike the latter, there is a coupling $\lam$ of negative dimension, which allows us to have currents $(T_n, \Theta_{n-2})$ for which both components have both holomorphic and antiholomorphic derivatives.

\section{Discussion} \label{Sec5}

The perspective in this paper has been the following. Given any quantum field theory, the $T \b T$ flow equation, 
\be \label{DiffEIvD}
\frac{\d \mL(\lam)}{\d \lam} = - 4 \( T_{z z} ^{\lam}\,  T_{\b z \b z}^{\lam} - (T_{z \b z}^{\lam})^2\)~
\ee
gives a one-parameter family of theories, with $\lam=0$ corresponding to the original theory. This equation serves as the definition of the $T\b T$ deformed theory. We interpret all quantities in this equation as  renormalized and UV finite. From this equation, it is simple to find the $\lam$ dependence of certain quantities, such as the energy spectrum and the $S$-matrix, and (perhaps) difficult to find certain other quantities, such as correlation functions. Our perspective has been that we will use the flow equation (or, more precisely, the $S$-matrix which follows from it) to find the renormalized Lagrangian of the $T\b T$ deformed theory. Once we have the renormalized Lagrangian we can, in principle, compute all other quantities  using standard QFT perturbation theory. Of course, computing with the Lagrangian, or defining the theory through the Lagrangian, is not necessarily optimal. Indeed, for the $T\b T$ theories, the simplicity of the $S$-matrix and of the flow equation (\ref{DiffEIvD}) suggests that these are  ultimately better definitions and starting points for computations. Nevertheless it is, at the very least, conceptually useful to make contact with the  Lagrangian. 

In Sec.~\ref{Massive} we computed, to one loop order, the renormalized Lagrangian of the $T \b T$ deformation of a free massive scalar. An interesting question is what the Lagrangian looks like to all orders; the simplicity of the $S$-matrix gives reason to be optimistic that there is a way of writing the Lagrangian so that it too looks simple. Our study has been of the $T\b T$ deformation of an integrable theory. It would be of interest to repeat the calculation for the $T\b T$ deformation of an non-integrable theory. It is not obvious that in this case it is possible to construct a renormalized Lagrangian satisfying the necessary properties. 

\section*{Acknowledgements} \noindent We thank  
A.~Cavagli\`a, S.~Dubovsky,  D.~Gross, S.~Komatsu, P.~Kraus, J.~Maldacena, S.~Minwalla, D.~Simmons-Duffin, and H.~Verlinde for helpful discussions.  
The work of VR is supported  by NSF grant NSF PHY-1911298. VR thanks the Aspen Center for Physics, NSF grant PHY-1607611, for hospitality while this work was being completed. The work of MS is supported by the Binational Science Foundation (grant No. 2016186) and by the ”Quantum Universe” I-CORE program of the Israel Planning and Budgeting Committee (grant No. 1937/12).

\appendix

\section{One-loop Integrals} \label{AppendixG}
In this appendix we evaluate some  one-loop integrals that arise in the computation of the $S$-matrix of two-dimensional massive theories. 
\subsection*{Preliminaries}
We start with the finite integral, 
\be \label{D1}
\int  \frac{d^2 k}{(2\pi)^2} \frac{1}{(k^2 - m^2)^2} = i \int  \frac{d^2 k_E}{(2\pi)^2} \frac{1}{(k_E^2 +m^2)^{2}} =  \frac{ i }{ 4\pi  m^2}~,
\ee
where we Wick rotated to Euclidean signature, $k_0 = i k_0^{E}$ and $k^2 = - k_E^2$. Next, consider the divergent integral, which we regulate by placing a hard cutoff,
\be
 \int  \frac{d^2 k}{(2\pi)^2} \frac{k^2}{(k^2 - m^2)^{2}} =i\int_0^{ \Lambda}  \frac{d k_E}{2\pi} \frac{-k_E^3}{(k_E^2 +m^2)^{2}}=- \frac{i}{4\pi} \( - \frac{\Lam^2}{\Lam^2 + m^2} + \log\( \frac{\Lam^2+m^2}{m^2}\) \)~.
\ee
The right-hand side is exact; we may simplify it by dropping terms that go to zero as $\Lambda$ goes to infinity. We will simplify it further, and drop all terms that are finite. We will do the same for all other divergent integrals, keeping only $\log$ terms and $\Lam^2$ terms (in fact, there is little point in keeping the $\Lambda^2$ terms, but we will do so anyway). Thus, we write the above integral as, 
\be
 \int  \frac{d^2 k}{(2\pi)^2} \frac{k^2}{(k^2 - m^2)^{2}} = \frac{i}{4\pi} \log\( \frac{m^2}{\Lam^2}\)~.
\ee
Using Lorentz symmetry we can evaluate the following integral, 
 \be
\int  \frac{d^2 k}{(2\pi)^2} \frac{k_{\mu} k_{\nu}}{(k^2 - m^2)^{2}} = \frac{i}{8\pi}\, \eta_{\mu \nu}  \log\( \frac{m^2}{\Lam^2}\)~.
\ee
We will need  two additional integrals. The first is, 
\be \label{I4}
\int  \frac{d^2 k}{(2\pi)^2} \frac{k_{\mu} k_{\nu} k_{\al} k_{\beta}}{(k^2 - m^2)^{2}} = \frac{i}{32\pi}\(\eta_{\mu \nu} \eta_{\al \beta} + \eta_{\mu \al} \eta_{\nu \beta} + \eta_{\mu \beta} \eta_{\nu \al}\)\,  \(\Lam^2+ 2 m^2 \log\( \frac{m^2}{\Lam^2}\)  \)~,
\ee
and the second is, 
\be \label{C6}
\int \frac{d^2 k }{(2\pi)^2} \frac{1}{p^2 - m^2} = \frac{i}{4\pi} \log\( \frac{m^2}{\Lam^2}\)~.
\ee
 
\subsection*{One-loop integrals}
We now turn to the one-loop integrals that appear in the calculation of scattering amplitudes. 
We start with the finite one-loop integral, 
\be \label{L1}
L_1(p^2) = \int \frac{d^2 k}{(2\pi)^2} \frac{1}{k^2 - m^2} \frac{1}{(p-k)^2 - m^2}~.
 \ee
Introducing Feynman parameters, 
this becomes, 
\bea
L_1(p^2)
&=&\int_0^1 d x\int \frac{d^2 k}{(2\pi)^2} \frac{1}{\[ (k-p(1-x))^2 +p^2 x(1-x) - m^2\]^2 }   \\
&=& \int_0^1 d x\int \frac{d^2 k}{(2\pi)^2}  \frac{1}{\[ k^2 +p^2 x(1-x) - m^2\]^2 }~,
\eea
where in the second equality we shifted the momentum $k\rightarrow  k+p(1-x)$. Using (\ref{D1}) to evaluate the $k$ integral gives, 
\be \label{G7}
L_1(p^2)=\frac{i}{4\pi} \int_0^1 d x\,\frac{1}{\(m^2 - x(1-x) p^2\)}~. 
\ee
Through a partial fraction decomposition, one can establish that, 
\be \label{D10}
\int_0^1dx \frac{1}{A-x(1-x)} =\frac{2}{\sqrt{1-4A}}\log\frac{ \sqrt{1- 4A}-1}{\sqrt{1- 4A}+1}~.
\ee
Thus our integral $L_1(p^2)$ becomes,
\be
L_1(p^2) =  \frac{i}{2\pi p^2  \sqrt{1- 4 m^2/p^2}} \log \frac{\sqrt{1- 4 m^2/p^2}-1}{  \sqrt{1- 4 m^2/p^2}+1}=\frac{i}{2\pi s \sqrt{-t/s}} \log \frac{\sqrt{-t/s}-1}{  \sqrt{-t/s}+1}~,
\ee
where in the second equality we used the Mandelstam variables $s= p^2$ and $t = 4 m^2 - p^2$. 
Let us write this in terms of $\theta$, which we take to be positive. The Mandelstam variables are,
\be
s= p^2 = p_+ p_- = 2 m^2(1+\cosh \theta) = 4 m^2 \cosh^2 \frac{\theta}{2}~, \ \ \ \ \Rightarrow t= - 4m^2 \sinh^2 \frac{\theta}{2}~.
\ee
and so the expression appearing in $L_1(p^2)$ is, 
\be
\frac{\sqrt{-t/s}-1}{  \sqrt{-t/s}+1} = - e^{- \theta}~, \ \ \ \ \ \sqrt{- s t} = 2 m^2 \sinh \theta~.
\ee
As a result we have that, 
\be \label{Slog}
\frac{1}{\sqrt{- s t}} \log\frac{\sqrt{-t/s}-1}{  \sqrt{-t/s}+1}   = \frac{ ( i \pi\, - \theta)}{2 m^2 \sinh \theta}~,
\ee
where we used that $s$ is really $s+i\eps$, in order to pick the correct sheet. Hence, our integral is,
\be \label{G12}
L_1(s) =  \frac{i}{4\pi}\,\frac{ ( i \pi\, - \theta)}{m^2 \sinh \theta}~.
\ee
Next, we consider the  one-loop integral $L_{\mu \nu}$ defined by, 
\be
L_{\mu \nu} =\frac{1}{4} \int \frac{d^2 k}{(2\pi)^2} \frac{k_{\mu} }{k^2 - m^2} \frac{(p_{\nu} - k_{\nu})}{(p-k)^2 - m^2}~.
\ee
As before, we introduce Feynman parameters, perform the $k$ integral, and then perform the $x$ integral. This gives, 
\be \label{L++v2}
L_{++} = \frac{i}{16\pi} \frac{p_+^2}{s}\( - 1 +\frac{ ( i \pi\, - \theta)}{\sinh \theta}\)~, \ \ \ \ \ L_{- + } = \frac{i }{16\pi}\left(-\log \left(\frac{m^2}{\Lambda ^2}\right)+ (i \pi - \theta) \frac{\cosh\theta}{\sinh \theta}\right)~.
\ee
In evaluating $L_{- +}$, to perform the $x$ integral we used, 
\be \label{G34}
\int_0^1 dx\, \log\( A- x(1-x)\)  = \log A -2 + \frac{1}{2} \int_0^1 dx\, \frac{ 4 A -1}{A - x(1-x)}~,
\ee
where the integral on the right is just the integral (\ref{D10})  that appeared earlier. 
This identity, as well as similar ones, are easily established through integration by parts. 
Finally, we look at the integral $L_{\mu \nu \al \beta}$, 
\be
L_{\mu \nu \al \beta}=\frac{1}{16} \int \frac{d^2 k}{(2\pi)^2} \frac{k_{\mu} k_{\nu} }{k^2 - m^2} \frac{(p_{\al} - k_{\al}) (p_{\beta} - k_{\beta})}{(p-k)^2 - m^2}~. 
\ee
Evaluating gives, 
 \bea  \nn
L_{++++} &=&  \frac{i}{64 \pi}\frac{p_+^4 m^2}{s^2} \(   - \frac{4}{3} - \frac{\cosh \theta}{3} + \frac{ ( i \pi\, - \theta)}{ \sinh \theta}\) \\[3pt] \nn
L_{-+++} &=& \frac{i\,  p_+^2}{64 \pi}\frac{ m^2}{3 s}\Big( 1 - 2\cosh\theta + 3(i\pi - \theta)\frac{\cosh \theta}{\sinh \theta}\Big)\\[3pt] \nn
L_{- - + + } &=&\frac{i m^2}{64\pi} \left(\frac{\Lambda ^2}{m^2}-2 \cosh\theta  \log \frac{m^2}{\Lambda ^2}+(i \pi -\theta)\frac{ \cosh (2\theta)}{\sinh \theta}\right)\\[3pt] \label{L++++}
L_{+ - + -} &=&\frac{i m^2}{64\pi} \left(\frac{\Lambda^2}{m^2} - \frac{1}{3} \cosh \theta + 2 \log \frac{m^2}{\Lambda ^2} + \frac{(i \pi - \theta)}{\sinh \theta} \right)~.
\eea

\section{Two-dimensional Massless Integrals} \label{IntEx}

In this Appendix we record some two-dimensional integrals. 
In two dimensions, we use the notation $\[z\]^{a}$, which denotes,
\be
\frac{1}{\[z\]^a} \equiv \frac{1}{z^a {\b z}^{\b a}} = \frac{{\b z}^{a - \b a}}{|z|^{2 a}} = \frac{z^{\b a - a}}{|z|^{2 \b a}} = \frac{(-1)^{a - \b a}}{\[-z\]^a}~,
\ee
where the difference  between $a$ and $\b a$ is an integer, $a - \b a = n$. 
\subsection*{Two-point Integral}
The two-point integral is given by, 
\be \label{2d2pt}
\int d^2 x_3 \frac{1}{\[z_{13}\]^{\al} \[z_{32}\]^{\beta}} = \pi (-1)^{\gamma - \b \gamma} C(\al) C(\beta) C(\gamma)~\frac{1}{\[z_{12}\]^{\al + \beta -1}}~,
\ee
where $\gamma = 2 - \al - \beta$ and  $z_{i j} \equiv z_i - z_j$ and,\be \label{C}
C(\al) = \frac{\G(1- \b \al)}{\G(\al)} = (-1)^{\al - \b \al} \frac{\G(1 - \al)}{\G(\b \al)}~, 
\ee
where the second equality is a consequence of the difference in $\al$ and $\b \al$ being an integer.
This integral can be easily obtained through Fourier transform for the case of $\al = \b \al $ and $\beta = \b \beta$, combined with differentiation for the other cases.  A few two-point integrals that we will need in particular are, 
\be\label{A4}
\int d^2 x_3 \frac{1}{z_{13}^2 z_{23}^{\beta} \b z_{23}^{\beta+2}} = - \pi \frac{\beta}{ \beta +1} \frac{1}{z_{12}^{\beta+1} \b z_{12}^{\beta+1}}~, 
\ee
\be
\int d^2 x_3 \frac{1}{ z_{23}^2 z_{13}^{\beta} \b z_{13}^{\beta}} = - \pi \frac{\beta}{ \beta-1} \frac{1}{z_{12}^{\beta+1}\b z_{12}^{\beta-1}}~, \ \  \ \ \ \ \ \ \int d^2 x_3 \frac{1}{z_{23}^2 } \frac{1}{z_{13}^{\beta} \b z_{13}^{\beta+1}}= -  \pi \frac{1}{z_{12}^{\beta+1} \b z_{12}^{\beta}}~.
\ee

\subsection*{Three-point Integral}

The three-point integral is given by,
\be
\int d^2 x_4 \frac{1}{\[z_{41}\]^{a} \[z_{42}\]^{b} \[z_{43}\]^{c}} = \frac{1}{\[z_{12}\]^{\frac{a+b}{2}}} \frac{1}{\[z_{13}\]^{\frac{a+b+2c-2}{2}}}\( A_{a, b, c}\,  f(\chi, \b \chi) + B_{a, b, c}\, g(\chi, \b \chi)   \)~,
\ee
where $ \chi = z_{12}/ z_{13}$ and, 
\bea
A_{a, b, c} &=& \pi \frac{\G(1- a- b)}{\G(\b a + \b b)}\frac{\G(1-c)}{\G(\b c)} \frac{\G(\b a + \b b + \b c-1)}{\G(2 - a - b -c )}\\
B_{a, b, c} &=& \pi \frac{\G(1- a)}{\G(\b a)} \frac{\G(1-\b b)}{\G(b)}\frac{\G(a+b-1)}{\G(2- \b a - \b b)}\\
f(\chi, \b \chi) &=& \chi^{\frac{a+b}{2}} {}_2 F_1(b, a+b+c-1, a+b, \chi)\,  \, \b \chi^{\frac{\b a+\b b}{2}} {}_2 F_1(\b b, \b a+\b b+\b c-1, \b a+\b b, \b \chi)\\
g(\chi, \b \chi)&=& \chi^{\frac{2- a -b}{2}}\, {}_2 F_1(c, 1-a, 2-a-b, \chi)\, \, \b \chi^{\frac{2- \b a -\b b}{2}}\, {}_2 F_1(\b c, 1-\b a, 2-\b a -\b b, \b \chi)~.
\eea
To get this result, we started with the expression for a conformal four-point integral and took one of the points to infinity.~\footnote{The conformal four-point integral is well known, as it is equivalent to the  integral definition of a  conformal partial wave, see e.g. \cite{Osborn:2012vt, Ferrara:1974ny}. An efficient way to evaluate the conformal partial wave integral is by noting that the partial wave is a sum of a conformal block and a shadow block (with coefficients that can be established by taking the $\chi\rightarrow 0$ limit). The two-dimensional conformal block is a product of two one-dimensional conformal blocks, which are trivial to establish. For more details, see for instance \cite{Rosenhaus:2018zqn, Gross:2017aos}.}
A particular three-point integral that we will need is, 
\be \label{3ptI}
\int d^2 x_4\, \frac{1}{z_{14}^2 z_{24}^2 \b z_{34}^4} =  \frac{2\pi }{3} \frac{1}{z_{12}^3} \( \frac{1}{\b z_{23}^{\, 3} } - \frac{1}{\b z_{13}^{\, 3} }\)~.
\ee

\subsection*{Double Integral}
A particular double integral that we will need is  \cite{Korchemsky}, 
\bml \nn
\hspace{-.3cm}\int d^2 x_0\, d^2 x_1\, \frac{\[z_{01}\]^{\al +\beta + \gamma -2}}{\[z_0\]^{\alpha} \[z_1\]^{\al} \[1-z_0\]^{\beta} \[1- z_1\]^{ \beta}}
 = \pi^2 \cos( \fdfrac{\pi}{2}(\small{ \al - \b \al + \beta - \b \beta + \gamma - \b \gamma}))\, 2^{\al + \b \al + \beta + \b \beta +\gamma + \b \gamma -4} \\
 C(\al) C(\beta) C(\gamma) 
  C(\fdfrac{3 - \al - \beta - \gamma}{2}) C(\fdfrac{2 - \al - \beta + \gamma}{2}) C(\fdfrac{2+ \al -\beta - \gamma}{2}) C( \fdfrac{2 - \al +\beta - \gamma}{2})~,
\end{multline}
where $C(\al)$ was defined in (\ref{C}).~\footnote{ An efficient way to evaluate such an integral is to add an auxiliary point, in such a way as to make the integral conformal. The resulting integral, which is similar to ones studied in\cite{Liu:2018jhs}, can then by evaluated through use of the Lorentzian inversion formula \cite{Caron-Huot:2017vep, Simmons-Duffin:2017nub}. }
Taking $\al =\b \beta =2, \b \al= \beta = 0$ and $\gamma = \b \gamma = -2 +\eps$ and sending $\eps$ to zero gives, 
\be \label{DI}
 \int d^2 x_3 d^2 x_4 \frac{1}{z_{13}^2 z_{34}^2 z_{41}^2}\frac{1}{ \b z_{23}^2 \b z_{34}^2 \b z_{42}^2} = \frac{\pi^2}{2} \frac{1}{z_{12}^4 \b z_{12}^{\, 4}}~.
\ee

\bibliographystyle{utphys}

\begin{thebibliography}{99}


\bibitem{SZ}
F.~A. Smirnov and A.~B. Zamolodchikov, ``{On space of integrable quantum field
  theories},'' \href{http://dx.doi.org/10.1016/j.nuclphysb.2016.12.014}{{\em
  Nucl. Phys.} {\bf B915} (2017)  363--383},
\href{http://arxiv.org/abs/1608.05499}{{\tt arXiv:1608.05499 [hep-th]}}.

\bibitem{ZZ}
A.~B. Zamolodchikov and A.~B. Zamolodchikov, ``{Factorized s Matrices in
  Two-Dimensions as the Exact Solutions of Certain Relativistic Quantum Field
  Models},'' \href{http://dx.doi.org/10.1016/0003-4916(79)90391-9}{{\em Annals
  Phys.} {\bf 120} (1979)  253--291}.
[,559(1978)].

\bibitem{Tateo16}
A.~Cavagli\`a, S.~Negro, I.~M. Sz\'ecs\'enyi, and R.~Tateo, ``{$T
  \bar{T}$-deformed 2D Quantum Field Theories},''
  \href{http://dx.doi.org/10.1007/JHEP10(2016)112}{{\em JHEP} {\bf 10} (2016)
  112},
\href{http://arxiv.org/abs/1608.05534}{{\tt arXiv:1608.05534 [hep-th]}}.

\bibitem{Bonelli:2018kik}
G.~Bonelli, N.~Doroud, and M.~Zhu, ``{$T \bar{T}$-deformations in closed
  form},'' \href{http://dx.doi.org/10.1007/JHEP06(2018)149}{{\em JHEP} {\bf 06}
  (2018)  149},
\href{http://arxiv.org/abs/1804.10967}{{\tt arXiv:1804.10967 [hep-th]}}.

\bibitem{Dubovsky:2012wk}
S.~Dubovsky, R.~Flauger, and V.~Gorbenko, ``{Solving the Simplest Theory of
  Quantum Gravity},'' \href{http://dx.doi.org/10.1007/JHEP09(2012)133}{{\em
  JHEP} {\bf 09} (2012)  133},
\href{http://arxiv.org/abs/1205.6805}{{\tt arXiv:1205.6805 [hep-th]}}.

\bibitem{Caselle:2013dra}
M.~Caselle, D.~Fioravanti, F.~Gliozzi, and R.~Tateo, ``{Quantisation of the
  effective string with TBA},''
  \href{http://dx.doi.org/10.1007/JHEP07(2013)071}{{\em JHEP} {\bf 07} (2013)
  071},
\href{http://arxiv.org/abs/1305.1278}{{\tt arXiv:1305.1278 [hep-th]}}.

\bibitem{Dubovsky:2013ira}
S.~Dubovsky, V.~Gorbenko, and M.~Mirbabayi, ``{Natural Tuning: Towards A Proof
  of Concept},'' \href{http://dx.doi.org/10.1007/JHEP09(2013)045}{{\em JHEP}
  {\bf 09} (2013)  045},
\href{http://arxiv.org/abs/1305.6939}{{\tt arXiv:1305.6939 [hep-th]}}.

\bibitem{Z}
A.~B. Zamolodchikov, ``{Expectation value of composite field T anti-T in
  two-dimensional quantum field theory},''
\href{http://arxiv.org/abs/hep-th/0401146}{{\tt arXiv:hep-th/0401146
  [hep-th]}}.

\bibitem{Dubovsky:2017cnj}
S.~Dubovsky, V.~Gorbenko, and M.~Mirbabayi, ``{Asymptotic fragility, near
  AdS$_{2}$ holography and $ T\overline{T} $},''
  \href{http://dx.doi.org/10.1007/JHEP09(2017)136}{{\em JHEP} {\bf 09} (2017)
  136},
\href{http://arxiv.org/abs/1706.06604}{{\tt arXiv:1706.06604 [hep-th]}}.

\bibitem{Dubovsky:2018bmo}
S.~Dubovsky, V.~Gorbenko, and G.~Hernández-Chifflet, ``{$ T\overline{T} $
  partition function from topological gravity},''
  \href{http://dx.doi.org/10.1007/JHEP09(2018)158}{{\em JHEP} {\bf 09} (2018)
  158},
\href{http://arxiv.org/abs/1805.07386}{{\tt arXiv:1805.07386 [hep-th]}}.

\bibitem{Cardy}
J.~Cardy, ``{The $ T\overline{T} $ deformation of quantum field theory as
  random geometry},'' \href{http://dx.doi.org/10.1007/JHEP10(2018)186}{{\em
  JHEP} {\bf 10} (2018)  186},
\href{http://arxiv.org/abs/1801.06895}{{\tt arXiv:1801.06895 [hep-th]}}.

\bibitem{Conti:2018tca}
R.~Conti, S.~Negro, and R.~Tateo, ``{The $ \mathrm{T}\overline{\mathrm{T}} $
  perturbation and its geometric interpretation},''
  \href{http://dx.doi.org/10.1007/JHEP02(2019)085}{{\em JHEP} {\bf 02} (2019)
  085},
\href{http://arxiv.org/abs/1809.09593}{{\tt arXiv:1809.09593 [hep-th]}}.

\bibitem{Conti:2018jho}
R.~Conti, L.~Iannella, S.~Negro, and R.~Tateo, ``{Generalised Born-Infeld
  models, Lax operators and the $ \mathrm{T}\overline{\mathrm{T}} $
  perturbation},'' \href{http://dx.doi.org/10.1007/JHEP11(2018)007}{{\em JHEP}
  {\bf 11} (2018)  007},
\href{http://arxiv.org/abs/1806.11515}{{\tt arXiv:1806.11515 [hep-th]}}.

\bibitem{Herman}
L.~McGough, M.~Mezei, and H.~Verlinde, ``{Moving the CFT into the bulk with $
  T\overline{T} $},'' \href{http://dx.doi.org/10.1007/JHEP04(2018)010}{{\em
  JHEP} {\bf 04} (2018)  010},
\href{http://arxiv.org/abs/1611.03470}{{\tt arXiv:1611.03470 [hep-th]}}.

\bibitem{Guica:2019nzm}
M.~Guica and R.~Monten, ``{$T\bar T$ and the mirage of a bulk cutoff},''
\href{http://arxiv.org/abs/1906.11251}{{\tt arXiv:1906.11251 [hep-th]}}.

\bibitem{Hartman}
T.~Hartman, J.~Kruthoff, E.~Shaghoulian, and A.~Tajdini, ``{Holography at
  finite cutoff with a $T^2$ deformation},''
  \href{http://dx.doi.org/10.1007/JHEP03(2019)004}{{\em JHEP} {\bf 03} (2019)
  004},
\href{http://arxiv.org/abs/1807.11401}{{\tt arXiv:1807.11401 [hep-th]}}.

\bibitem{Taylor:2018xcy}
M.~Taylor, ``{TT deformations in general dimensions},''
\href{http://arxiv.org/abs/1805.10287}{{\tt arXiv:1805.10287 [hep-th]}}.

\bibitem{Kutasov}
A.~Giveon, N.~Itzhaki, and D.~Kutasov, ``{$ \mathrm{T}\overline{\mathrm{T}} $
  and LST},'' \href{http://dx.doi.org/10.1007/JHEP07(2017)122}{{\em JHEP} {\bf
  07} (2017)  122},
\href{http://arxiv.org/abs/1701.05576}{{\tt arXiv:1701.05576 [hep-th]}}.

\bibitem{Giribet:2017imm}
G.~Giribet, ``{$T\bar{T}$-deformations, AdS/CFT and correlation functions},''
  \href{http://dx.doi.org/10.1007/JHEP02(2018)114}{{\em JHEP} {\bf 02} (2018)
  114},
\href{http://arxiv.org/abs/1711.02716}{{\tt arXiv:1711.02716 [hep-th]}}.

\bibitem{Eva}
V.~Gorbenko, E.~Silverstein, and G.~Torroba, ``{dS/dS and $ T\overline{T} $},''
  \href{http://dx.doi.org/10.1007/JHEP03(2019)085}{{\em JHEP} {\bf 03} (2019)
  085},
\href{http://arxiv.org/abs/1811.07965}{{\tt arXiv:1811.07965 [hep-th]}}.

\bibitem{Datta:2018thy}
S.~Datta and Y.~Jiang, ``{$T\bar{T}$ deformed partition functions},''
  \href{http://dx.doi.org/10.1007/JHEP08(2018)106}{{\em JHEP} {\bf 08} (2018)
  106},
\href{http://arxiv.org/abs/1806.07426}{{\tt arXiv:1806.07426 [hep-th]}}.

\bibitem{Jiang:2019hxb}
Y.~Jiang, ``{Lectures on solvable irrelevant deformations of 2d quantum field
  theory},''
\href{http://arxiv.org/abs/1904.13376}{{\tt arXiv:1904.13376 [hep-th]}}.

\bibitem{Aharony:2018bad}
O.~Aharony, S.~Datta, A.~Giveon, Y.~Jiang, and D.~Kutasov, ``{Modular
  invariance and uniqueness of $T\bar{T}$ deformed CFT},''
  \href{http://dx.doi.org/10.1007/JHEP01(2019)086}{{\em JHEP} {\bf 01} (2019)
  086},
\href{http://arxiv.org/abs/1808.02492}{{\tt arXiv:1808.02492 [hep-th]}}.

\bibitem{Arefeva:1974bk}
I.~Arefeva and V.~Korepin, ``{Scattering in two-dimensional model with
  Lagrangian (1/gamma) ((d(mu)u)**2/2 + m**2 cos(u-1))},''
{\em Pisma Zh. Eksp. Teor. Fiz.} {\bf 20} (1974)  680.

\bibitem{Townsend}
L.~Mezincescu and P.~K. Townsend, ``{Anyons from Strings},''
  \href{http://dx.doi.org/10.1103/PhysRevLett.105.191601}{{\em Phys. Rev.
  Lett.} {\bf 105} (2010)  191601},
\href{http://arxiv.org/abs/1008.2334}{{\tt arXiv:1008.2334 [hep-th]}}.

\bibitem{Conkey:2016qju}
P.~Conkey and S.~Dubovsky, ``{Four Loop Scattering in the Nambu-Goto Theory},''
  \href{http://dx.doi.org/10.1007/JHEP05(2016)071}{{\em JHEP} {\bf 05} (2016)
  071},
\href{http://arxiv.org/abs/1603.00719}{{\tt arXiv:1603.00719 [hep-th]}}.

\bibitem{Brown:1979pq}
L.~S. Brown, ``{Dimensional Regularization of Composite Operators in Scalar
  Field Theory},''
\href{http://dx.doi.org/10.1016/0003-4916(80)90377-2}{{\em Annals Phys.} {\bf
  126} (1980)  135}.

\bibitem{Brown}
L.~S. Brown, {\em Quantum Field Theory}.
\newblock Cambridge University Press, 1992.

\bibitem{Kraus}
P.~Kraus, J.~Liu, and D.~Marolf, ``{Cutoff AdS$_{3}$ versus the $ T\overline{T}
  $ deformation},'' \href{http://dx.doi.org/10.1007/JHEP07(2018)027}{{\em JHEP}
  {\bf 07} (2018)  027},
\href{http://arxiv.org/abs/1801.02714}{{\tt arXiv:1801.02714 [hep-th]}}.

\bibitem{Ofer}
O.~Aharony and T.~Vaknin, ``{The TT* deformation at large central charge},''
  \href{http://dx.doi.org/10.1007/JHEP05(2018)166}{{\em JHEP} {\bf 05} (2018)
  166},
\href{http://arxiv.org/abs/1803.00100}{{\tt arXiv:1803.00100 [hep-th]}}.

\bibitem{Guica:2019vnb}
M.~Guica, ``{On correlation functions in $J\bar T$-deformed CFTs},''
  \href{http://dx.doi.org/10.1088/1751-8121/ab0ef3}{{\em J. Phys.} {\bf A52}
  (2019) no.~18, 184003},
\href{http://arxiv.org/abs/1902.01434}{{\tt arXiv:1902.01434 [hep-th]}}.

\bibitem{Cardy:2019qao}
J.~Cardy, ``{$T\overline T$ deformation of correlation functions},''
\href{http://arxiv.org/abs/1907.03394}{{\tt arXiv:1907.03394 [hep-th]}}.

\bibitem{Baggio:2018rpv}
M.~Baggio, A.~Sfondrini, G.~Tartaglino-Mazzucchelli, and H.~Walsh, ``{On $
  T\overline{T} $ deformations and supersymmetry},''
  \href{http://dx.doi.org/10.1007/JHEP06(2019)063}{{\em JHEP} {\bf 06} (2019)
  063},
\href{http://arxiv.org/abs/1811.00533}{{\tt arXiv:1811.00533 [hep-th]}}.

\bibitem{Chang:2018dge}
C.-K. Chang, C.~Ferko, and S.~Sethi, ``{Supersymmetry and $ T\overline{T} $
  deformations},'' \href{http://dx.doi.org/10.1007/JHEP04(2019)131}{{\em JHEP}
  {\bf 04} (2019)  131},
\href{http://arxiv.org/abs/1811.01895}{{\tt arXiv:1811.01895 [hep-th]}}.

\bibitem{Guica:2017lia}
M.~Guica, ``{An integrable Lorentz-breaking deformation of two-dimensional
  CFTs},'' \href{http://dx.doi.org/10.21468/SciPostPhys.5.5.048}{{\em SciPost
  Phys.} {\bf 5} (2018) no.~5, 048},
\href{http://arxiv.org/abs/1710.08415}{{\tt arXiv:1710.08415 [hep-th]}}.

\bibitem{Mark}
B.~Le~Floch and M.~Mezei, ``{Solving a family of $T\bar{T}$-like theories},''
\href{http://arxiv.org/abs/1903.07606}{{\tt arXiv:1903.07606 [hep-th]}}.

\bibitem{Frolov:2019nrr}
S.~Frolov, ``{TTbar deformation and the light-cone gauge},''
\href{http://arxiv.org/abs/1905.07946}{{\tt arXiv:1905.07946 [hep-th]}}.

\bibitem{Jiang:2019tcq}
Y.~Jiang, ``{Expectation value of $\mathrm{T}\overline{\mathrm{T}}$ operator in
  curved spacetimes},''
\href{http://arxiv.org/abs/1903.07561}{{\tt arXiv:1903.07561 [hep-th]}}.

\bibitem{David}
D.~J. Gross, J.~Kruthoff, A.~Rolph, and E.~Shaghoulian, ``{$T\overline{T}$ in
  AdS$_2$ and Quantum Mechanics},''
\href{http://arxiv.org/abs/1907.04873}{{\tt arXiv:1907.04873 [hep-th]}}.

\bibitem{LeFloch:2019wlf}
B.~Le~Floch and M.~Mezei, ``{KdV charges in $T\bar{T}$ theories and new models
  with super-Hagedorn behavior},''
\href{http://arxiv.org/abs/1907.02516}{{\tt arXiv:1907.02516 [hep-th]}}.

\bibitem{Sfondrini:2019smd}
A.~Sfondrini and S.~J. van Tongeren, ``{$T\bar{T}$ deformations as TsT
  transformations},''
\href{http://arxiv.org/abs/1908.09299}{{\tt arXiv:1908.09299 [hep-th]}}.

\bibitem{KR}
S.~Komatsu and V.~Rosenhaus, ``{Comments on $T_{s+1} \bar T_{s+1}$
  deformations},'' {\em unpublished}  .

\bibitem{Bajnok}
L.~Samaj and Z.~Bajnok, {\em Introduction to the Statistical Physics of
  Integrable Many-Body Systems}.
\newblock Cambridge University Press, 2013.

\bibitem{Dorey:1996gd}
P.~Dorey, ``{Exact S matrices},'' in {\em {Conformal field theories and
  integrable models. Proceedings, Eotvos Graduate Course, Budapest, Hungary,
  August 13-18, 1996}}, pp.~85--125.
\newblock 1996.
\newblock
\href{http://arxiv.org/abs/hep-th/9810026}{{\tt arXiv:hep-th/9810026
  [hep-th]}}.
\newblock

\bibitem{Gabai:2018tmm}
B.~Gabai, D.~Mazáč, A.~Shieber, P.~Vieira, and Y.~Zhou, ``{No Particle
  Production in Two Dimensions: Recursion Relations and Multi-Regge Limit},''
  \href{http://dx.doi.org/10.1007/JHEP02(2019)094}{{\em JHEP} {\bf 02} (2019)
  094},
\href{http://arxiv.org/abs/1803.03578}{{\tt arXiv:1803.03578 [hep-th]}}.

\bibitem{Osborn:2012vt}
H.~Osborn, ``{Conformal Blocks for Arbitrary Spins in Two Dimensions},''
  \href{http://dx.doi.org/10.1016/j.physletb.2012.09.045}{{\em Phys. Lett.}
  {\bf B718} (2012)  169--172},
\href{http://arxiv.org/abs/1205.1941}{{\tt arXiv:1205.1941 [hep-th]}}.

\bibitem{Ferrara:1974ny}
S.~Ferrara, R.~Gatto, and A.~F. Grillo, ``{Properties of Partial Wave
  Amplitudes in Conformal Invariant Field Theories},''
\href{http://dx.doi.org/10.1007/BF02769009}{{\em Nuovo Cim.} {\bf A26} (1975)
  226}.

\bibitem{Rosenhaus:2018zqn}
V.~Rosenhaus, ``{Multipoint Conformal Blocks in the Comb Channel},''
  \href{http://dx.doi.org/10.1007/JHEP02(2019)142}{{\em JHEP} {\bf 02} (2019)
  142},
\href{http://arxiv.org/abs/1810.03244}{{\tt arXiv:1810.03244 [hep-th]}}.

\bibitem{Gross:2017aos}
D.~J. Gross and V.~Rosenhaus, ``{All point correlation functions in SYK},''
  \href{http://dx.doi.org/10.1007/JHEP12(2017)148}{{\em JHEP} {\bf 12} (2017)
  148},
\href{http://arxiv.org/abs/1710.08113}{{\tt arXiv:1710.08113 [hep-th]}}.

\bibitem{Korchemsky}
G.~P. Korchemsky, ``{Conformal bootstrap for the BFKL pomeron},''
  \href{http://dx.doi.org/10.1016/S0550-3213(99)00185-6}{{\em Nucl. Phys.} {\bf
  B550} (1999)  397--423},
\href{http://arxiv.org/abs/hep-ph/9711277}{{\tt arXiv:hep-ph/9711277
  [hep-ph]}}.

\bibitem{Liu:2018jhs}
J.~Liu, E.~Perlmutter, V.~Rosenhaus, and D.~Simmons-Duffin, ``{$d$-dimensional
  SYK, AdS Loops, and $6j$ Symbols},''
  \href{http://dx.doi.org/10.1007/JHEP03(2019)052}{{\em JHEP} {\bf 03} (2019)
  052},
\href{http://arxiv.org/abs/1808.00612}{{\tt arXiv:1808.00612 [hep-th]}}.

\bibitem{Caron-Huot:2017vep}
S.~Caron-Huot, ``{Analyticity in Spin in Conformal Theories},''
  \href{http://dx.doi.org/10.1007/JHEP09(2017)078}{{\em JHEP} {\bf 09} (2017)
  078},
\href{http://arxiv.org/abs/1703.00278}{{\tt arXiv:1703.00278 [hep-th]}}.

\bibitem{Simmons-Duffin:2017nub}
D.~Simmons-Duffin, D.~Stanford, and E.~Witten, ``{A spacetime derivation of the
  Lorentzian OPE inversion formula},''
  \href{http://dx.doi.org/10.1007/JHEP07(2018)085}{{\em JHEP} {\bf 07} (2018)
  085},
\href{http://arxiv.org/abs/1711.03816}{{\tt arXiv:1711.03816 [hep-th]}}.

\end{thebibliography}

\end{document}